\documentclass[preprint,journal]{vgtc}       

\onlineid{1104}

\vgtccategory{Research}

\vgtcpapertype{application/design study}

\title{\systemname: A Visual Analytics Approach for Understanding the Dual Frontiers of Science and Technology}

\author{ 
    \authororcid{Yifang Wang}{0000-0001-6267-9440}, 
    \authororcid{Yifan Qian}{0000-0002-3914-1981},
    \authororcid{Xiaoyu Qi}{0009-0003-0292-6368},
    \authororcid{Nan Cao*}{0000-0003-1316-7515},
    \authororcid{Dashun Wang*}{0000-0002-7054-2206} 
} 

\authorfooter{
\item
    Yifang Wang, Yifan Qian, and Dashun Wang are with The Center for Science of Science and Innovation, Northwestern University. E-mail: \{yifang.wang, yifan.qian1, dashun.wang\}@kellogg.northwestern.edu.
\item
    Xiaoyu Qi and Nan Cao are with Intelligent Big Data Visualization Lab, Tongji University. E-mail: \{qixiaoyu, nan.cao\}@tongji.edu.cn. 
\item 
    * Dashun Wang and Nan Cao are the co-corresponding authors. 
}

\abstract{
Science has long been viewed as a key driver of economic growth and rising standards of living. 
Knowledge about how scientific advances support marketplace inventions is therefore essential for understanding the role of science in propelling real-world applications and technological progress. 
The increasing availability of large-scale datasets tracing scientific publications and patented inventions and the complex interactions among them offers us new opportunities to explore the evolving dual frontiers of science and technology at an unprecedented level of scale and detail. 
However, we lack suitable visual analytics approaches to analyze such complex interactions effectively. 
Here we introduce \systemname, an interactive visual analysis system for researchers, research institutions, and policymakers to explore the complex linkages between science and technology, and to identify critical innovations, inventors, and potential partners. 
The system first identifies important associations between scientific papers and patented inventions through a set of statistical measures introduced by our experts from the field of the Science of Science. 
A series of visualization views are then used to present these associations in the data context. 
In particular, we introduce the \interplayview to visualize patterns and insights derived from the data, helping users effectively navigate citation relationships between papers and patents. 
This visualization thereby helps them identify the origins of technical inventions and the impact of scientific research. 
We evaluate the system through two case studies with experts followed by expert interviews. 
We further engage a premier research institution to test-run the system, helping its institution leaders to extract new insights for innovation. 
Through both the case studies and the engagement project, we find that our system not only meets our original goals of design, allowing users to better identify the sources of technical inventions and to understand the broad impact of scientific research; 
it also goes beyond these purposes to enable an array of new applications for researchers and research institutions, ranging from identifying untapped innovation potential within an institution to forging new collaboration opportunities between science and industry. 
}


\graphicspath{{figs/}{figures/}{pictures/}{images/}{./}} 

\usepackage{tabu}                      
\usepackage{booktabs}                  
\usepackage{lipsum}                    
\usepackage{mwe}                       

\usepackage{mathptmx}                  

\usepackage{balance}
\usepackage{enumitem}
\usepackage{setspace}
\usepackage{xspace}
\usepackage{microtype}                 
\PassOptionsToPackage{warn}{textcomp}  
\usepackage{textcomp}                  
\usepackage{mathptmx}                  
\usepackage{times}                     
\usepackage{cite}                      
\usepackage{tabu}                      
\usepackage{booktabs}                  
\usepackage{color}
\usepackage[dvipsnames,svgnames]{xcolor}
\usepackage[most]{tcolorbox}
\usepackage{graphicx}

\definecolor{variable-background}{RGB}{244, 244, 245}
\definecolor{variable-border}{RGB}{136, 136, 136}
\definecolor{variable-text}{RGB}{175, 175, 175}
\newtcbox{\variable}[1][]
  {on line, 
   colframe= variable-border, 
   colback = variable-background, 
   arc = 2pt, 
   boxsep = 0pt, 
   left = 2pt, right = 2pt, top = 2pt, bottom = 2pt, 
   boxrule = 1pt, 
  }

\newcommand{\systemname}{\textit{InnovationInsights}\xspace} 
\newcommand{\researcheroverviewview}{\textit{Researcher Overview View}\xspace}
\newcommand{\researcherstatisticsview}{\textit{Researcher Statistics View}\xspace}
\newcommand{\innovationview}{\textit{Innovation View}\xspace}
\newcommand{\technologyinspectionview}{\textit{Technology Inspection View}\xspace}
\newcommand{\scienceinspectionview}{\textit{Science Inspection View}\xspace}

\newcommand{\interplayview}{\textit{Interplay Graph}\xspace}
\newcommand{\fieldtimeline}{\textit{Field Timeline}\xspace}
\newcommand{\papermatrix}{\textit{Paper Matrix}\xspace}
\newcommand{\patenttreemap}{\textit{Patent Icicle Plot}\xspace}
\newcommand{\citationflow}{\textit{Citation Flow}\xspace}

\newcommand{\etal}{et al.}
\newcommand{\ie}{i.e.}
\newcommand{\eg}{e.g.}
\newcommand{\ea}{$E_{A}$\xspace} 
\newcommand{\eb}{$E_{B}$\xspace} 
\newcommand{\ec}{$E_{C}$\xspace} 
\newcommand{\ed}{$E_{D}$\xspace} 
\newcommand{\newea}{$P_{A}$\xspace} 
\newcommand{\neweb}{$P_{B}$\xspace} 
\newcommand{\newec}{$P_{C}$\xspace} 
\newcommand{\newed}{$P_{D}$\xspace} 
\newcommand{\newee}{$P_{E}$\xspace} 
\newcommand{\newef}{$P_{F}$\xspace} 
\newcommand{\tasktypeone}{Researcher Identification\xspace} 
\newcommand{\tasktypetwo}{Interplay Exploration\xspace} 
\newcommand{\tasktypethree}{Invention Prediction\xspace} 
\newcommand{\tasktypeoneshort}{\textbf{I}} 
\newcommand{\tasktypetwoshort}{\textbf{E}} 
\newcommand{\tasktypethreeshort}{\textbf{P}} 
\newcommand{\universityname}{the university\xspace}

\definecolor{cameraready}{RGB}{237, 85, 106} 
\newcommand{\cameraready}[1]{\textcolor{black}{#1}} 
\newcommand{\secondround}[1]{\textcolor{black}{#1}} 

\usepackage{amsmath}
\usepackage{amsfonts}
\usepackage{float}

\begin{document}
\begin{spacing}{0.97} 
\firstsection{Introduction}
\label{sec:01_introduction}
\maketitle 

Science is central to improving the human condition~\cite{bush1990science, jones2021science}. 
Not only has science long been recognized as the engine for long-run economic growth and prosperity, but also it has been essential to creating critical solutions to confront emergent threats to humanity, from climate change to the COVID-19 pandemic.
While scientific research propels both fundamental understanding and practical applications~\cite{fortunato2018science,wang2021science,ahmadpoor2017dual, yin2022public}, there has been a lack of visual analytics approaches to explore the complex linkages (\ie, the dual frontiers) between scientific advances and technical inventions. 
Here we introduce \systemname, which represents an initial step toward filling this crucial gap.

A better understanding of the dual frontiers of science and technology informs a diverse range of stakeholders, from researchers and research institutions to policymakers and private companies, helping them identify gaps and opportunities for innovation while facilitating more rapid and effective knowledge transfer. 
For example, research institutions such as universities aim to discover untapped innovation potentials and forge new collaboration and partnership opportunities between science and industry. 
Companies seek to stay abreast of the latest scientific breakthroughs to drive the creation of new applications. 

The availability of large-scale datasets~\cite{patentsview, liang2022systematic, marx2020reliance} tracing scientific publications and patented inventions and the complex interactions among them has created new opportunities to tackle this research question. 
Here we build on the Science of Science literature (SciSci~\cite{fortunato2018science, wang2021science}), which provides descriptive insights into the connections between science and technology~\cite{ahmadpoor2017dual, yin2022public, liang2022systematic}. 
While SciSci has furthered our understanding of the uses of science both within and outside of science, 
it also highlights the numerous challenges to interactively explore the complex interactions among multiple entities, from inventors and inventions to scientists and their publications. 
Meanwhile, existing studies in the visualization community have primarily focused on papers~\cite{narechania2021vitality, li2019galex} or patents~\cite{kutz2004examining, koch2010iterative, ankam2012exploring} separately, rarely examining their interconnections. 

Here we hypothesize that visual analytics may provide an effective means to meet these new analytical demands. 
Designing such a system requires us to overcome several major challenges: 
(1) The complexity of the data, which encompasses networks, multi-dimensional attributes, hierarchical structures, and temporal features, poses visualization challenges for effectively navigating these complex interactions; 
(2) The multi-dimensional nature of the entities and complex linkages among them, coupled with multiple levels of granularity, pose analytical challenges for quantitative measures of the science-technology interface. 
(3) The massive amounts of underlying data, including papers, paper citations, patents, patent citations, and paper-to-patent citations, pose scalability challenges for the system. 
(4) Given the long pathways in knowledge transfer, we need new predictive models to accompany visual approaches to identify innovation gaps and opportunities.

To tackle these challenges, we first characterize the problem domain and define a set of statistical measures in collaboration with our domain experts. 
We then develop a prediction model to estimate the extent to which scientific advances may propel future technological applications, allowing us to systematically identify a list of innovators whose work holds considerable potential for commercialization. 
Based on these measures and models, 
we develop a visualization system with multi-dimensional views into the complex relationships between science and technology. We introduce the \interplayview, which enables interactive exploration of the dual frontiers of science and technology in a scalable manner. 
Our contributions are summarized as follows: 
\begin{itemize}[leftmargin=10pt,topsep=1pt,itemsep=1px]
    \item We formulate the domain of visual analysis of dual frontiers of science and technology and propose a novel design to visualize the complex interactions between science and technology. 
    \vspace{-0.2cm}
    \item We design and develop \systemname, which to the best of our knowledge, is the first visual analysis system to explore 
    rich interactions between upstream scientific research and downstream technological development. 
    \vspace{-0.2cm}
    \item We conduct comprehensive evaluations, including case studies and expert interviews, to demonstrate the effectiveness of our system. Moreover, we engage with a premier research university and its institutional leaders as a trial run for the developed visual analytics systems, helping key stakeholders uncover new insights for gaps and untapped potentials for innovation in real-world settings.
\end{itemize}
Overall, the system we developed serves as an initial but crucial step toward using visual analytics to bridge the ivory tower and the real world, helping amplify the real-world impact of scientific research while significantly advancing the R\&D success of research institutions. 
\section{Related Work}
\label{sec:02_RelatedWork} 

The related literature spans three domains, including the science of science, visualization of scientific data, and graph and tree visualization. 

\subsection{The Science of Science} 
\label{sec:02_RelatedWork_SciSciAnalytics} 
The increasing availability of large-scale data tracing nearly all phases of scientific production and use has fueled the emergence of an interdisciplinary field, SciSci, to explore opportunities and premises to accelerate scientific discoveries. 
Despite the rapid progress in this field,  the bulk of the literature has focused on the impact of science within science, ranging from the unfolding of scientific careers~\cite{liu2021understanding} to the scientific impact of papers~\cite{wang2013quantifying}, to scientific collaborations~\cite{wu2019large}. 

More recently, studies have made initial attempts to quantify the broad impact of science~\cite{ahmadpoor2017dual, marx2020reliance, yin2021coevolution, yin2022public}, aiming to better understand the interface between science and various facets of human society, from policy-making~\cite{yin2021coevolution} to public perception~\cite{yin2022public}. 
In particular, the interaction between science and technology is a critical area of focus, with studies testing theories that emphasize the connections between patents and prior scientific advances. 
Ahmadpoor and Jones~\cite{ahmadpoor2017dual} conducted the first systematic analyses into \textit{the dual frontier of science and technology}, finding that advances that sit directly at the science-technology interface are significantly more impactful within their respective domains.
Marx and Fuegi~\cite{marx2020reliance,marx2022reliance} created a large-scale dataset, ``Reliance on Science,'' to trace citations from patents to papers. 
Yin~\etal\cite{yin2022public} introduced an index to quantify the extent to which papers from a scientific field are consumed by patents. 
Cao~\etal~\cite{cao2023breaking} specifically targeted the HCI community and studied the impact of HCI papers on the industry. 
Overall, these efforts contribute to a data-driven understanding of the dual frontiers. 
Yet, the statistical nature of these studies highlights the lack of visualization approaches to studying this important problem. 
This is especially true given the substantial challenges in identifying patterns and extracting insights hidden beneath complex data structures. 

In this paper, we aim to provide an interactive visual analytics approach to help experts analyze the complex interactions between science and technology more systematically and effectively.

\subsection{Scientific Data Visualization} 
\label{sec:02_RelatedWork_VAInSciSci} 
Large-scale scientific data represent a familiar domain in the visualization community, with several comprehensive reviews written on the subject~\cite{federico2016survey,chen2003mapping,borner2010atlas}. 
Here we examine pertinent studies from these surveys and recent advances in the visualization community, focusing on two main categories: (1) visualization for scientific data, and (2) the science of science within the visualization community. 

\textbf{Visualization for scientific data.} 
Many visualization techniques have been developed to reveal insights from scientific databases, 
such as dynamic and heterogeneous networks
~\cite{dork2012pivotpaths, ye2022visatlas}, 
sequences and time series~\cite{wu2015egoslider, wang2021seek}, 
multi-dimension measures~\cite{nobre2018juniper}, 
and texts~\cite{dou2013hierarchicaltopics, gonzalez2023landscape}. 
These studies introduce general visualization techniques for specific data structures and use scientific datasets as illustrative examples. 
Other research focuses on developing visualization systems to streamline literature queries (\eg, VitaLITy~\cite{narechania2021vitality} and \secondround{VisualBib~\cite{dattolo2022authoring}}
), scientific discoveries (\eg, VIStory~\cite{dong2019vistory} and Galex~\cite{li2019galex}), and academic evaluations (\eg, SD$^2$~\cite{guo2022sd}). 
These studies only use data within science (\ie, papers). 

Similar to papers, patent data are also of a textual, network, and temporal nature. 
Studies using patent data cover several themes: 
patent document classification~\cite{kim2008visualization, ankam2012exploring} and information retrieval~\cite{boyack2000analysis, koch2010iterative}, 
knowledge discovery using patent citation, semantic, or agent (\eg, inventor and organization) networks~\cite{boyack2000analysis, kim2008visualization, windhager2015concept}, 
and technology evolution and emerging technology exploration~\cite{ankam2012exploring, windhager2015concept}. 
In addition, a few studies provide a comprehensive analysis by visualizing multifaceted patent data. 
\secondround{PatViz~\cite{koch2010iterative} summarizes patent search results using various factors (\eg, patent categories), but it does not focus on the relationship between papers and patents.} 
\secondround{DIVA~\cite{morris2002diva}, on the other hand, creates network linkages between papers and patents using keyword similarity. 
However, it fails to provide deep insights between science and technology due to the absence of paper-patent citations.} 

\textbf{Science of science within the visualization community.} 
A growing number of studies have explored scientific data within the visualization community to reflect on community development. 
These studies involve the creation of specific datasets like Vispubdata~\cite{isenberg2016vispubdata}, VIS30K~\cite{chen2021vis30k}, and VisImages~\cite{deng2022visimages}, and the development of  platforms such as VIS Author Profiles~\cite{latif2018vis} and VISPubComPAS~\cite{wang2019vispubcompas}. 
Additionally, many works statistically analyze the community's development, focusing on aspects such as authors~\cite{hao2022thirty}, topics~\cite{hao2022thirty}, genders~\cite{sarvghad2022scientometric, tovanich2021gender}, collaborations~\cite{sarvghad2022scientometric}, peer reviews~\cite{wu2022defence}, and so on. 

Overall, these studies have focused on papers or patents separately, ignoring the complex interactions between them. 
Our work combines multiple data sources for papers and patents to systematically study the dual frontiers of science and technology. 

\subsection{Graph and Tree Visualization}  
\label{sec:02_RelatedWork_NetworkVisualization} 
Graph~\cite{beck2017taxonomy, wang2021interactive, deng2022multilevel, abdelaal2022comparative} and tree~\cite{pandey2021state, wang2021m2lens, li2022gotreescape} visualization have been studied extensively in the visualization community. 
Here we discuss the most relevant techniques for paper-patent citation networks, including bipartite graphs, compound graphs, and tree visualization. 

Patent-paper citations may be represented as a bipartite graph problem. 
Sun \etal~\cite{sun2018effect} proposed a technique, bicluster-based seriation, to reduce edge crossings in a bipartite graph. 
Chan \etal~\cite{chan2018v} used the minimum description length (MDL) principle to aggregate bipartite relations for scalability. 
However, they disregard complex node attributes, such as temporal and hierarchical structures, rendering them unsuitable for our scenario. 
The compound graph technique, often used for large-scale networks to group nodes for scalability, is another relevant method~\cite{beck2017taxonomy}. 
Our work also uses this technique to visualize the large-scale citation graph by grouping paper and patent nodes. 

Tree visualization displays hierarchical node connections~\cite{pandey2021state}. 
Various tree types use the width of the link to represent the flow quantity between the parent and child nodes, such as decision trees (\eg, BaobabView~\cite{van2011baobabview} and TreePOD~\cite{muhlbacher2017treepod}) and flow maps~\cite{phan2005flow, buchin2011flow}. 
However, most methods focus on static graphs and neglect the hierarchical aspect of nodes. 
They are insufficient for our needs, as we require temporal dimensions and detailed exploration across various levels of detail.  

We, therefore, propose a scalable node-link representation that summarizes the citation between papers and patents, overcoming various domain-specific constraints. 
\section{System Design}
\label{sec:03_Background} 
In this section, we summarize the analysis goals and design tasks and introduce the system overview.

\subsection{Analysis Goals} 
\label{sec:03_Background_ProblemFormulation} 
Over the past two years, we have been working closely with leaders from a premier US research university to understand and predict a university's innovation landscape and potential. 
We first collected private data from various organizations in the university, including (1) Technology Transfer Office (TTO) on invention disclosures and outcomes, licensing, and startups; (2) HR Office on faculty roster with demographic data (e.g., name, gender, rank, and department); and (3) Sponsored Research Office on grant applications and \secondround{their outcomes (granted or rejected)}. 
We then linked these data with global innovation databases on science and technology, capturing publications, patents, and how these publications are cited in patents as prior art, spanning all scientific fields and patenting domains. 
After a massive data cleaning and linkage, we used statistical methods for data analysis.   
\secondround{We then presented our findings to leaders from seven top US research universities (\ie, two public land-grant, three public non-land-grant, and two private universities), four R\&D-based companies and venture firms, and four science funders.}
During these interactions with leaders in science, industry, and government, we saw great interest in identifying untapped innovation potential in research institutions across a wide range of stakeholders. 
Given the novelty of the research question and the diverse array of stakeholders it informs, 
we quickly realized the need for a visual system, which would be crucial for efficient analysis and communication among multiple stakeholders. 

To this end, we initiated an interdisciplinary collaboration with domain experts from various fields to design a visual analytics system nine months ago.   
Two experts, \ea and \eb, are from the TTO of our partner university. 
\ea is a senior director of invention management who helps faculty convert their scientific output into patents. 
\eb is an analyst providing data-driven support. 
The other two experts, \ec and \ed, are SciSci researchers. 
\ec is a well-known professor in the field of SciSci. 
\ed is a postdoc who focuses on statistical measurements of university innovations. 
They all aim to understand how scientific research influences technology development and to find promising scientific directions and researchers with high innovation potential. 
To achieve this objective, they seek to explore the interplay between science and technology comprehensively. 
Existing methods rely on statistical analysis via ad-hoc analysis procedures, lacking an integrated system to help identify key innovators and research topics. 
The experts particularly identified three critical analysis goals:

\begin{enumerate}[topsep=1pt,itemsep=2px,partopsep=2pt,parsep=2pt]
\item[{\bf (I)}] {\bf \tasktypeone.} Identifying researchers and research ideas with high potential for practical applications and commercialization opportunities will help research institutions to better support researchers with a greater level of equity and efficiency. 

\item[{\bf (E)}] {\bf \tasktypetwo.} Exploring the interplay between science and technology is another goal frequently mentioned by experts. 
The exploration requires not only the extraction of relational patterns (\eg, paper-patent citations) but also the interpretation of these connections in the context of knowledge transfer. 

\item[{\bf (P)}] {\bf \tasktypethree.} Being able to predict the future likelihood of patentability and the commercialization potential of scientific research and technological inventions is a capability that is highly desired by our domain experts. 
\end{enumerate}

\subsection{Visualization Design Tasks} 
\label{sec:03_Background_AnalyticalTasks} 
\secondround{Given the goals above}, we devised the design tasks by following \secondround{the expert-focused design study methodology~\cite{sedlmair2012design}}. 
To this end, literature reviews 
guided by our experts, requirement analysis via expert interviews, as well as brainstorming sessions with both visualization and SciSci experts, were conducted iteratively. The design tasks for each analysis goal are outlined below: 
\\
\\
\centerline{\bf Tasks for \tasktypeone (\tasktypeoneshort)}
\begin{enumerate}[topsep=1pt,itemsep=2px,partopsep=2pt,parsep=2pt]
\item[{\bf \tasktypeoneshort1:}] {\bf Provide Researcher Overview.} The visual analysis system should provide an overview of researchers to help analysts select individuals of interest for further investigation by aligning researchers based on their research profiles.

\item[{\bf \tasktypeoneshort2:}] {\bf Create Productivity Portrait.} The system should illustrate the detailed characteristics of individual researchers to help the analysts identify talent and potential. Two types of information are of particular relevance: (1) demographic information--such as gender, age, and job rank--for equity and inclusion; (2) productivity and impact measurements for research output.\\
\end{enumerate}

\centerline{\textit{\bf Tasks for \tasktypetwo (\tasktypetwoshort)}}
\begin{enumerate}[topsep=1pt,itemsep=2px,partopsep=2pt,parsep=2pt]
\item[{\bf \tasktypetwoshort1:}]{\bf Inspect the Interplay.} The visualization design should be scalable enough to display large-scale linkages between scientific papers and technical inventions at both individual and field levels intuitively. The design should also clearly reveal the papers that are cited heavily by patents and highlight their characteristics to help \secondround{analysts} understand the key factors in knowledge transfer. 


\item[{\bf \tasktypetwoshort2:}]{\bf Reveal Temporal Trends.} The design should reveal the temporal changes in research topics and the corresponding technical developments to help \secondround{analysts} understand the evolving frontiers of science and technology.

\item[{\bf \tasktypetwoshort3:}]{\bf Reason with Contextual Information.} Beyond showing the interplay based on explicit linkages between patents and papers, the visualization should capture contextual information to help \secondround{analysts} uncover implicit linkages between research and technology. The contextual information can be captured by SciSci measures of papers, fields, and the assignee information of patents, which is crucial for identifying important research or estimating the commercialization potential of an invention.\\
\end{enumerate}

\centerline{{\bf Tasks for \tasktypethree(\tasktypethreeshort)}}

\begin{enumerate}[topsep=1pt,itemsep=2px,partopsep=2pt,parsep=2pt]
\item[{\bf \tasktypethreeshort1:}]{\bf Identify Untapped Innovation Potential.} Our system should uncover untapped innovation potential in researchers, research topics, and scientific fields. 
This key capability building on the prediction model and intuitive visualization will significantly facilitate knowledge transfer for the latest scientific advances. 
\end{enumerate}

\begin{figure} [tb]
 \centering 
 \includegraphics[width=0.85\linewidth]{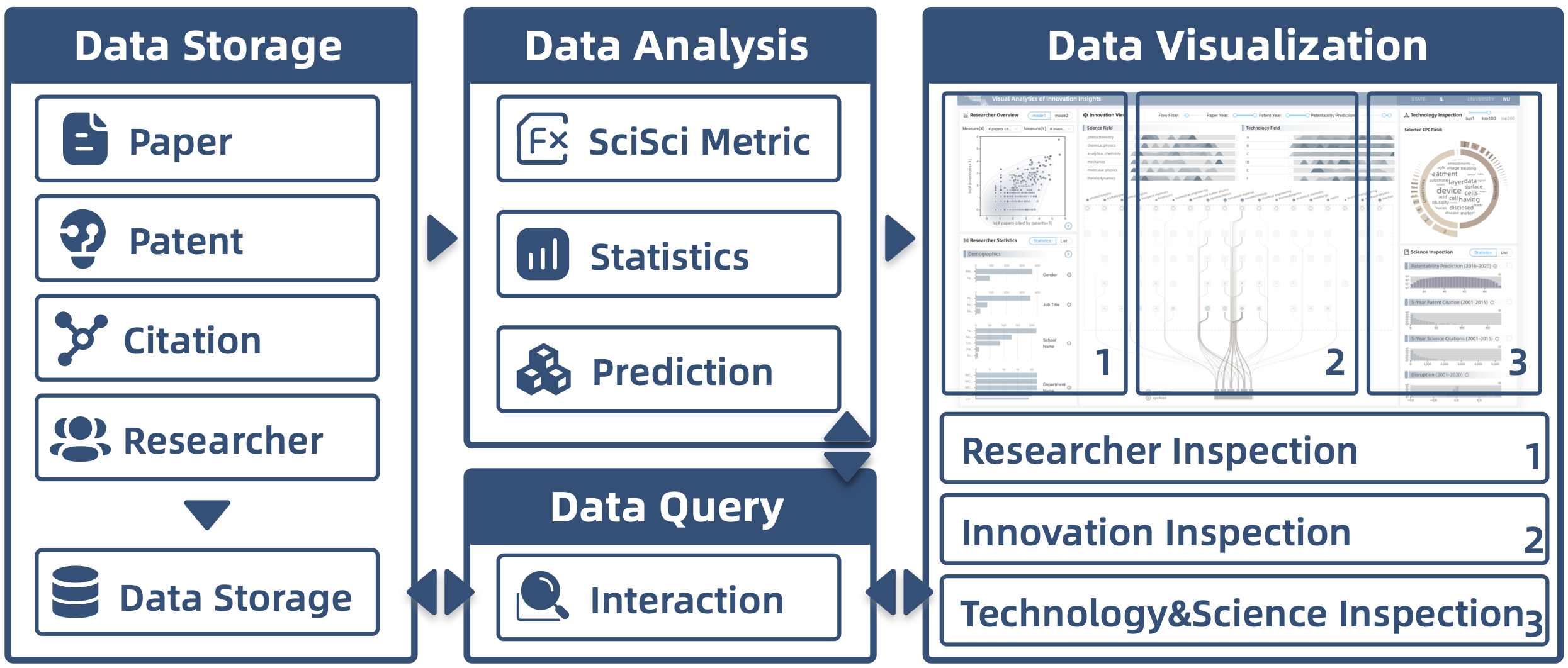}
 \vspace{-0.2cm}
 \caption{System overview. \systemname consists of a data storage module, a data analysis module, and a data visualization module.} 
 \vspace{-0.6cm}
 \label{fig:system-overview}
\end{figure}

\subsection{System Overview} 
\label{sec:03_Background_SystemOverview} 
Following the above analysis goals and design tasks, we design \systemname as an online system that consists of three modules (Fig.~\ref{fig:system-overview}): (1) the data storage module, (2) the data analysis module, and (3) the visualization module. 
The data storage module preprocesses data from multiple sources and stores it in a database. 
The data analysis module conducts a series of measurements on different entities (\eg, papers, patents, and researchers) and employs a prediction model to recommend papers with high patentability potential. 
The two modules form the backend of the system. 
Both historical and prediction data are fed into the data visualization module to display intuitive data insights. 



\section{Data Analysis}
\label{sec:04_DataAnalysis} 
The data analysis module is designed to calculate the contextual information for visual analysis and decision-making. Specifically, we consider two types of information: (1) the data facts about papers, patents, researchers, and assignees that are calculated based on a set of statistical metrics; (2) the potential of scientific research (i.e., a paper) to be transferred, which is estimated by a deep prediction model implemented based on a graph convolutional network (GCN). Before drilling into technique details, we will first introduce the data we use. 
\begin{figure*} [!htb]
 \centering 
 \vspace{-0.5cm}
 \includegraphics[width=0.95\linewidth]{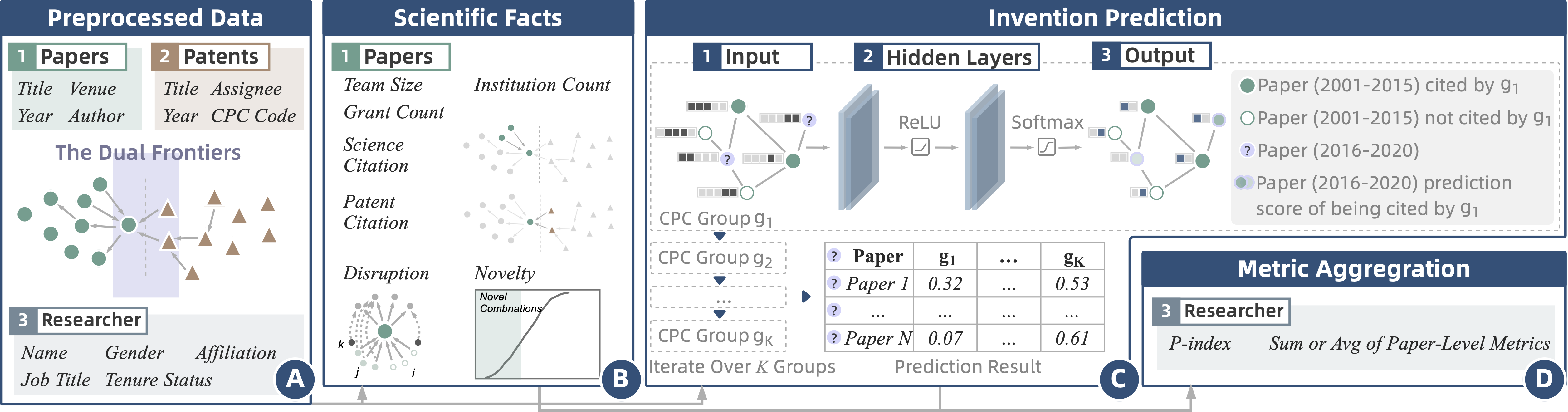}
 \vspace{-0.2cm} 
 \caption{The analytical process. 
 (A) We preprocess data into the network and multi-dimensional structures. 
 (B)-(C) Next, we construct SciSci metrics using scientific facts and predicted results at the paper level. 
 (D) Finally, we aggregate paper-level metrics to get researcher-level metrics. 
 } 
 \vspace{-0.6cm} 
 \label{fig:framework}
\end{figure*}

\subsection{Data Preprocessing} 
\label{sec:04_DataAnalysis_DataPreprocessing}  
Analyzing dual frontiers of science and technology needs to integrate data from various sources (Fig.~\ref{fig:framework}(A)), which are listed as follows:

\begin{itemize}[leftmargin=10pt,topsep=2pt,itemsep=1px]
    \item \textbf{Scientific Research Records}. We leverage the Microsoft Academic Graph (MAG) dataset ~\cite{wang2019review} to retrieve information about scientific research. The dataset consists of 270M research papers and their corresponding meta information, including the title, publication year, topic keywords, doi, author list, author affiliations, and citations. 
    \item \textbf{Technical Inventions}. We use the patent records collected in PatentsView~\cite{patentsview} to capture the technical inventions and reveal the development of technologies. This dataset contains over 7.9M patents filed through the United States Patent and Trademark Office (USPTO~\cite{uspto}). A subset of the most relevant patent attributes is carefully selected for analysis, including patent ID, title, application year, assignee name (\ie, the owner of the patent), and cooperative patent classification (CPC, \ie, the category of the patent~\cite{cpcscheme}). 
    The CPC category we used consists of three levels: section, subsection, and group (the lowest level). 
    Private data (\eg, invention disclosures and patents collected by research institutions) are also used. 
    \item \textbf{Science-Technology Linkages}.  To analyze the interplay between scientific research and the development of technology, we use data collected from ``Reliance on Science''~\cite{marx2020reliance, marx2022reliance}, which include more than 40M citations that record the details about how technical innovations (\ie, patents) cite research papers. 
    
    \item \textbf{Researcher Profiles.} This dataset provides demographic information (\eg, gender, rank, and affiliation) for each researcher, collected by research institutions \secondround{(\eg, gender and rank)} or automatically inferred by algorithms \secondround{(\eg, gender~\cite{karimi2016inferring})}. The data also contain each researcher's publication records, which are collected from public (\eg, MAG~\cite{wang2019review}) and private sources (\eg, university libraries). 
\end{itemize}

We extracted a subset of the data above \secondround{(supplementary material)} to demonstrate our idea of \secondround{analyzing} the dual frontiers of science and technology. 
This subset includes papers from a 20-year period (2001\secondround{--}2020). A group of researchers and patent assignees were also filtered out for analysis. In particular, the patent assignees were classified into three categories, i.e., university assignees, company assignees, and others, to provide additional context information. 

\subsection{Scientific Facts} 
\label{sec:04_DataAnalysis_SciSciMeasure} 
We analyze the data to capture the scientific facts for each research paper and each individual researcher based on a set of carefully defined metrics~\cite{lin2023sciscinet}. Specifically, given a research paper $\mathcal{P}$, the following metrics are designed to help \secondround{analysts} estimate the quality and impact of $\mathcal{P}$ in the context of knowledge transfer (Fig.~\ref{fig:framework}(B)), \secondround{which are calculated based on the entire dataset in Section~\ref{sec:04_DataAnalysis_DataPreprocessing}}: 
\begin{itemize}[leftmargin=10pt,topsep=1pt,itemsep=0px]
    \item \textit{Team Size}: the total number of co-authors of the paper. 
    \item \textit{Institution Count}: the total number of different affiliations regarding the co-authors of the paper. 
    \secondround{\item \textit{Grant Count}: the total number of grants sponsoring the research for the paper. Due to data availability, we focus primarily on grants from NSF and NIH as a demonstration.} 
    \item \textit{Science Citation}: the total number of citations the paper received within 5 years of publication. 
    \item \textit{Disruption}: the degree to which papers citing the focal paper $\mathcal{P}$ tend not to cite $\mathcal{P}$'s references~\cite{wu2019large}, which is formally defined as:
    \begin{equation}
        D = \frac{n_i - n_j}{n_i + n_j + n_k}
    \end{equation}
    where $n_i$ is the number of subsequent papers that only cite the focal paper, $n_j$ represents the number of subsequent papers that cite both the focal paper and its references, and $n_k$ represents the number of subsequent papers that only cite the references of the focal paper. 
    \item \textit{Novelty}: the extent to which the focal paper's combination of existing knowledge deviates from the norm among all journal pairs. We first calculate the z-score for each journal pair by comparing its observed frequency to the expected frequency in randomized citation graphs~\cite{uzzi2013atypical}. 
    The focal paper's novelty score is determined by the 10th percentile z-score of the journal pairs cited in its references. 
    \item \textit{Patent Citation}: the total number of patents that cite $\mathcal{P}$ within 5 years of its publication. This metric is used to measure $\mathcal{P}$'s impact on technical inventions.  
\end{itemize}

In addition, in order to estimate a researcher $\mathcal{R}$'s performance in scientific research and impact on technical inventions, we define the following metrics (Fig.~\ref{fig:framework}(D)): 
\begin{itemize}[leftmargin=10pt,topsep=1pt,itemsep=0px]
\item \textit{Paper Count}: the total number of papers that $\mathcal{R}$ has ever published.
\item \textit{Invention Count}: the total number of invention disclosures that $\mathcal{R}$ has ever disclosed to the university.
\item \textit{Scientific Citation}: the number of research papers that cite $\mathcal{R}$'s papers within 5 years of each paper’s publication. 
\item \textit{Number of Papers Cited by Patents}: the total number of papers that have been cited by at least one patent.  This metric measures $\mathcal{R}$'s impact on technical inventions.
\end{itemize} 

\begin{figure*} [!htb]
 \centering 
 \includegraphics[width=0.95\linewidth]{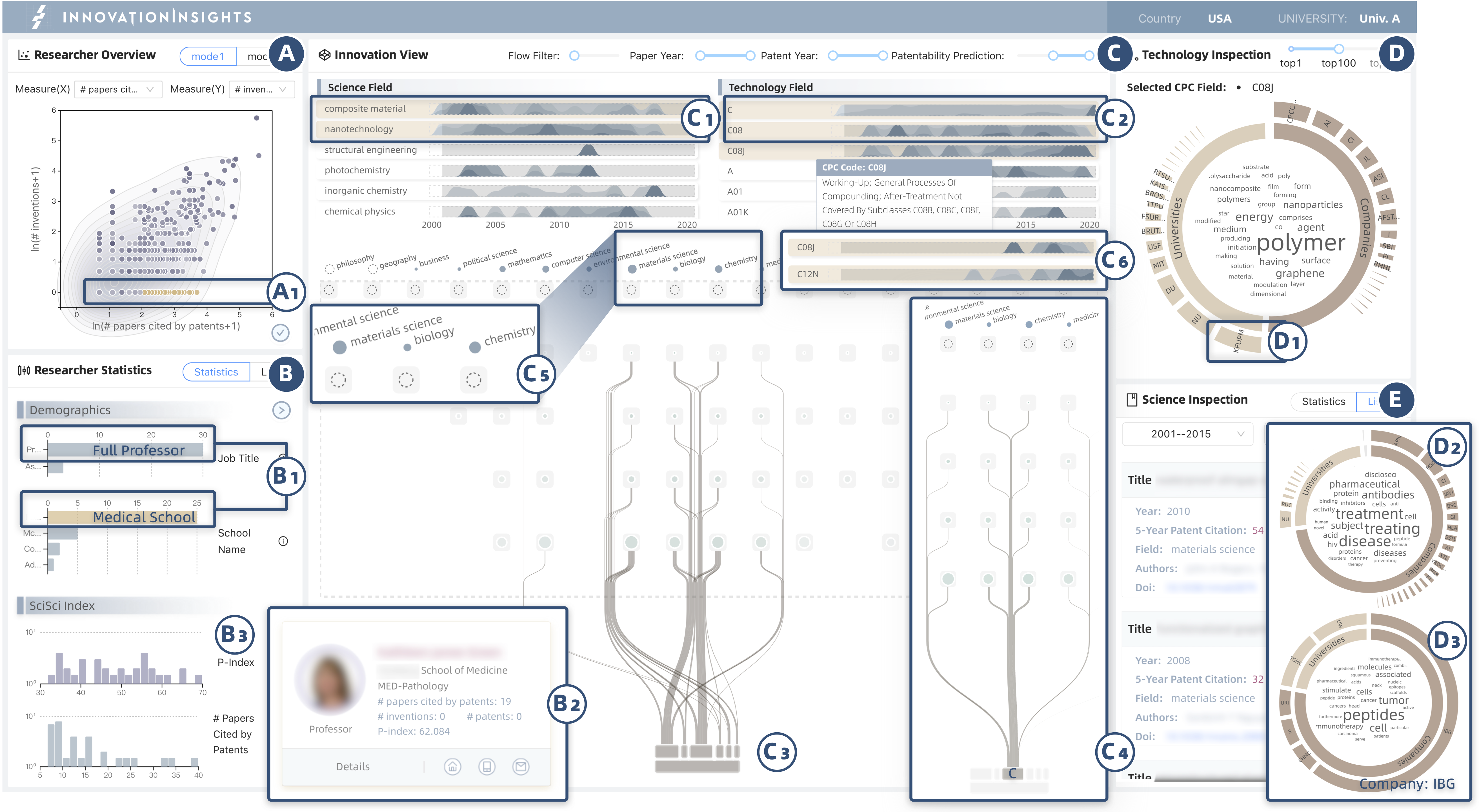}
 \vspace{-0.25cm} 
 \caption{The system UI of \systemname. 
The \researcheroverviewview (A) and \researcherstatisticsview (B) are for individual-level analysis. 
The \innovationview (C) shows the detailed interplay between science and technology. 
The \technologyinspectionview (D) and \scienceinspectionview (E) provide additional contextual information about patents and papers. 
 } 
 \vspace{-0.5cm} 
 \label{fig:system-ui}
\end{figure*}

\subsection{Invention Prediction}\label{sec:04_DataAnalysis_PredictionModel} 
The ability to estimate the potential of a paper for future inventions helps \secondround{analysts} to identify the next promising research topics as well as potential inventors. Here we introduce an invention prediction model to compute the probability that a research paper will spur future inventions in a given area. The model is designed based on the observation that when the knowledge obtained from a scientific research paper is used in a technical invention, the paper will be cited directly by the corresponding patent. Therefore, we use the citation links from patents to papers as a key feature to train a graph convolutional network (GCN) to help us estimate how likely (i.e., the probability) a paper $\mathcal{P}$ will be cited directly by patents from a specific technical area, as indicated by class labels (Fig.~\ref{fig:framework}(C)). We briefly review the architecture of GCN, followed by our implementation details below. 

\textbf{Graph Convolutional Network}. GCN is a crucial technique for deep learning on graph-based data~\cite{kipf2017semi}. 
It has transformed the field by providing a powerful way to analyze and model graph-structured data, which is common in many real-world applications such as social graphs, molecular graphs, and citation graphs~\cite{qian2021quantifying,qian2021geometric}. In our case, papers are connected by citations that form a citation graph, and thus GCN is naturally suited to our prediction task. In addition, compared to traditional deep learning models for grid-like data (e.g., images), GCN has several advantages. 
First, it can capture both local and global structures of the graph and also can handle graphs of varying sizes and structures. 
Second, it can be trained using both labeled and unlabeled data, i.e., in a semi-supervised learning way. By using unlabeled data, the model can learn more generalized representations of the data, which can improve its performance on the labeled data.

The input for GCN is a graph, in which each node is associated with an \secondround{$F$}-dimensional vector arranged as a row of the feature matrix $X \in \mathbb{R}^{N \times F}$. The adjacency matrix $A \in \mathbb{R}^{N \times N}$ represents the relationships between all nodes in the graph where the element $A_{ij}$ indicates the presence (1) or absence (0) of an edge between node $i$ and node $j$. The label matrix $Y \in \mathbb{R}^{N_{s} \times C}$, where $N_{s}$ is the number of nodes with labels in the graph and $C$ is the number of classes. The element $Y_{ij}$ is 1 if node $i$ belongs to class $j$, and 0 otherwise.

\textbf{Layer-wise Propagation Rule}. The standard GCN proposed by Kipf and Welling~\cite{kipf2017semi} takes two propagation layers to perform graph convolution operations on the input data. The first layer is defined as:
\begin{equation}
H = \mathrm{ReLU}(\widehat{A}XW^{0})
\label{eq:GCN_first_layer}
\end{equation}
where $W^{0}\in \mathbb{R}^{F \times B}$ is the weight matrix connecting the inputs and the first layer of the GCN. The graph is encoded in $\widehat{A}=\widetilde{D}^{-1/2}(A + I_{N})\widetilde{D}^{-1/2}$, where $I_{N}$ is the identity matrix, and $\widetilde{D}$ is a diagonal matrix with $\widetilde{D}_{ii} = 1 + \sum_{j}A_{ij}$. $\mathrm{ReLU}(\cdot)=\max(\cdot,0)$ is the activation function. $H\in \mathbb{R}^{N \times B}$ is the matrix of activations of the first layer.

The output layer is formally defined as follows:
\begin{equation}
Z = \mathrm{softmax}(\widehat{A}HW^{1})
\label{eq:GCN_second_layer}
\end{equation}
where $W^{1}\in \mathbb{R}^{B \times C}$ is the weight matrix connecting the first layer and output layer of the GCN. $\mathrm{softmax}(x)_{i} = \exp(x_{i})/\sum_{j} \exp(x_{j})$ where $x$ is a vector. $Z\in \mathbb{R}^{N \times C}$ is the output matrix of GCN where the element $Z_{ij}$ represents the probability of node $i$ belonging to class $j$.

\textbf{Loss Function}. GCN evaluates the cross-entropy error between the predicted class probabilities $Z$ and the true labels $Y$ for the nodes in the training set: 
\begin{equation}
    \mathcal{L} = -\sum_{l\in \mathbb{N}_{T}}\sum_{c=1}^{C} Y_{lc} \, \ln{Z_{lc}}
    \label{eq:cross_entropy_error}
\end{equation}
where $\mathbb{N}_{T}$ is the set of nodes in the training set.

\textbf{Implementation}. 
In our work, the input feature matrix $X$ is composed of paper title embeddings using SPECTER~\cite{specter_cohan_2020}, a popular API to generate embeddings for research papers. The input graph $A$ represents the citation graph between papers. Given a patent CPC category $g$, the label matrix $Y$ provides information on whether a paper is cited by a patent from CPC category $g$ within 5 years of publication. 
We selected a 5-year duration for the experiment because our experts prioritized recently published papers and aimed to determine if a paper would be cited by patents soon after its publication. 
We split the papers published between $2001$ and $2014$ into a training set (70\%) and a validation set (30\%), and use papers published in 2015 as the test set. For each paper published between $2016$ and $2020$, we predict its likelihood of being cited by patents in the CPC group $g$. We use the PyTorch Geometric implementation of GCN~\cite{Fey/Lenssen/2019} and follow the experimental setup proposed by Kipf and Welling~\cite{kipf2017semi}. Our model has $200$ epochs of training iterations, a learning rate of $0.01$, a dropout rate of $0.5$, and $16$ hidden units. The weights of the neural network ($W^{0}$ and $W^{1}$) are trained using gradient descent to minimize the loss~$\mathcal{L}$. To determine the likelihood of a paper $\mathcal{P}$ published between $2016$ and $2020$ being cited by patents in the CPC group $g$ within $5$ years of publication, we use the predicted probabilities in the softmax output matrix $Z$ obtained from the final model. In addition, to assess its relative importance within a specific range of papers (\eg, those within a research institution), we further convert its probability to a percentile, denoted as $\text{Patentability}_\mathcal{P}^g$, which is a scalar ranging from 0 to 100.

We apply the above prediction pipeline to the top $K$ patent CPC groups $g$ (denoted as $\mathbb{G}$) based on the number of patents citing our target papers. To assess a paper's overall patentability across different CPC groups, we compute its average likelihood of being cited by patent CPC groups $g$ in $\mathbb{G}$ (i.e., $\text{Patentability}_\mathcal{P}^g$), denoted as $\text{Patentability}_\mathcal{P}$. 

To evaluate a researcher's overall performance, we calculate the average $\text{Patentability}_\mathcal{P}$ of all their papers published between $2016$ and $2020$. This aggregated value is called the P-index. To the best of our knowledge, the P-index is the first index proposed in the SciSci literature that measures the extent to which a researcher's recent papers will be cited by patents in the future, acting as an indicator of a researcher's potential for commercial success. The higher the P-index, the higher the commercialization potential of the researcher. 




\section{Visualization}
\label{sec:05_VisualDesign} 
This section presents the visual design of \systemname. We introduce the user interface through a usage scenario followed by detailed descriptions of visualization views and corresponding interactions. 

\subsection{User Interface}
\label{sec:05_VisualDesign_UsageScenario} 
Fig.~\ref{fig:system-ui} illustrates the user interface of the proposed system, which \secondround{consists} of five coordinated views (Fig.~\ref{fig:system-ui}(A-E)). A user can start from the \researcheroverviewview (Fig.~\ref{fig:system-ui}(A)) to choose a group of researchers (\textbf{\tasktypeoneshort1}), whose scientific facts and profile information are summarized as the context in the \researcherstatisticsview(\textbf{\tasktypeoneshort2}, Fig.~\ref{fig:system-ui}(B)). The user can then filter to find interested researchers based on this contextual information. 
Two sets of horizon graphs are used to illustrate the trend of science and technologies (\textbf{\tasktypetwoshort2}) by displaying the changes in the numbers of the papers (Fig.~\ref{fig:system-ui}(C1)) and patents (Fig.~\ref{fig:system-ui}(C2)) in different fields. 
\secondround{The user} can brush a period of time to filter the papers or patents through these horizon graphs. Once the data (i.e., researchers, papers, and patents) are filtered out, an interplay graph (Fig.~\ref{fig:system-ui}(C3)) will visualize the citations between the selected patents and papers to reveal the interplay between science and technology (\textbf{\tasktypetwoshort1}). In this view, the user can interactively filter the citations via fields to learn the science-technology connections at different levels of detail. 
To give the user a better understanding of the connections, additional context information such as keywords for the selected patents (Fig.~\ref{fig:system-ui}(D)) and the list of the selected papers (Fig.~\ref{fig:system-ui}(E)) is also displayed (\textbf{\tasktypetwoshort3}). 

Despite the exploitative analysis above, the user can also investigate the potential inventors and inventions from multiple views based on the invention prediction results (\textbf{\tasktypethreeshort1}). For example, 
the user can observe the P-index scores of researchers in the \researcheroverviewview via the opacity of the circles. 
\secondround{The user} can also examine the researcher distribution (Fig.~\ref{fig:system-ui}(B3)) and rank the researchers based on P-index in the \researcherstatisticsview. 
In the \innovationview, the user can filter recent papers with high prediction scores using a slider (Fig.~\ref{fig:system-ui}(C)) and observe the prediction flow to identify the promising paper fields with high patentability potential. 
\secondround{The user} can also click a paper cell to check the details in the \scienceinspectionview.


\begin{figure} [tb]
 \centering 
 \includegraphics[width=0.92\linewidth]{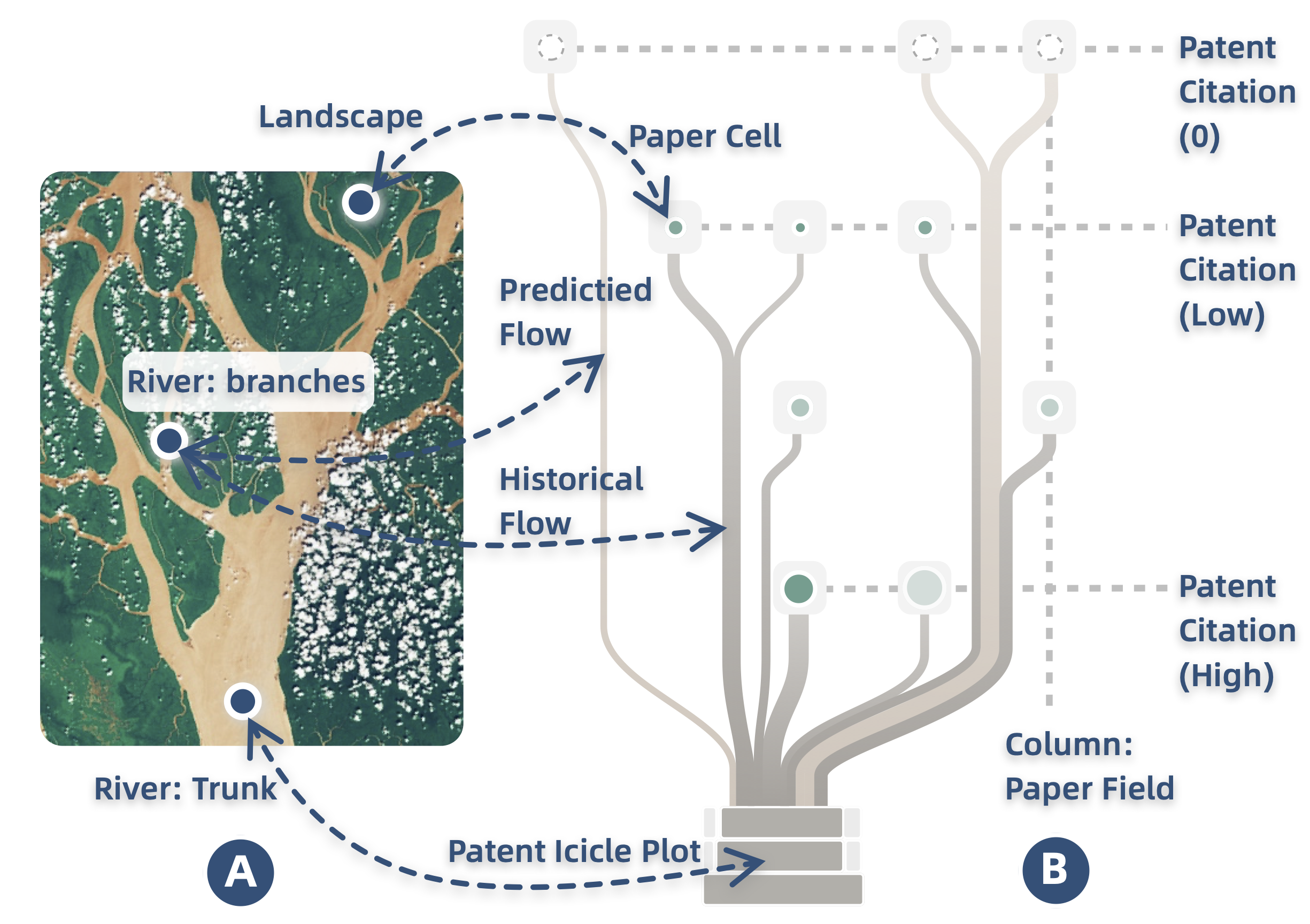}
 \vspace{-0.2cm}
 \caption{
    The design of the \interplayview is inspired by the river metaphor: 
    (A) the structure of the river; 
    (B) the river-like visual metaphor shows citation linkages between papers and patents with three components: Paper Matrix, Patent Icicle Plot, and Citation Flow. 
 } 
 \vspace{-0.65cm}
 \label{fig:visual-design}
\end{figure}

\subsection{Interplay Graph}
\label{sec:05_VisualDesign_Innovation_View_Interplay_Graph} 
The \interplayview (Fig.~\ref{fig:system-ui}(C3)) is the primary visualization component that enables users to explore the detailed citations from patents to papers and reveal the interplay between scientific research and technological inventions (\secondround{\textbf{\tasktypetwoshort1}, \textbf{\tasktypethreeshort1}}). It consists of three components: \papermatrix, \patenttreemap, and \citationflow, with a design that is inspired by a river metaphor (Fig.~\ref{fig:visual-design}). The \papermatrix symbolizes the knowledge landscape, and the \citationflow illustrates the diverse branches of knowledge that ultimately merge into the \patenttreemap, representing vast technological rivers. 

\textbf{\papermatrix}. 
The \papermatrix summarizes all the selected papers. \secondround{As the experts sought to explore the interplay at the level of the scientific field, we use} each column to represent a research field and each row to show a numerical citation range that indicates the extent to which the papers displayed in the row have been cited by patents. 
The citation number increases from the top to the bottom, i.e., the papers in the last row at the bottom of the view are those that are most cited by the patents. To deal with large-scale datasets, the \papermatrix supports interactive hierarchical aggregation via both columns and rows. In particular, the rows can be aggregated by directly merging the citation ranges; the columns, i.e., research fields, can be aggregated by following the research field hierarchy introduced in the MAG. 

A node in the \papermatrix indicates a collection of aggregated papers, which can be illustrated either by a circle or by a star glyph (Fig.~\ref{fig:case2}(B2)). The size of the circle indicates the number of papers in the collection, and the opacity represents the papers' averaged patent citation number. The star glyph summarizes the papers' statistical features (i.e., scientific facts in Section~\ref{sec:04_DataAnalysis_SciSciMeasure}). 

To reveal how broadly the papers in a research field $\mathcal{F}$ affect technologies across different areas, we compute a diversity score: 
\[
diversity = -\sum_{i=1}^{n}P(x_i)logP(x_i)
\]
where $i$ is a patent area. The score indicates the diversity of the area of the patents that cite the papers within $\mathcal{F}$. Intuitively, the larger the diversity score is, the larger $\mathcal{F}$'s influence will be. We use the size of the blue circle on top of each column (Fig.~\ref{fig:system-ui}(C5)) to show this score.

\textbf{\patenttreemap}. The \patenttreemap uses an upturned icicle plot~\cite{kruskal1983icicle} to summarize the patent CPC categories from a three-level hierarchy (Fig.~\ref{fig:system-ui}(C3)): section, subsection, and group~\cite{cpcscheme}. 
Each rectangle represents a category, with the length encoding the number of patents in that category. 
The patent fields are by default in alphabetical order, as required by our experts to query patent categories more efficiently. 
%

\textbf{\citationflow}. The \citationflow visualizes citation linkages from patents to papers. The flows start from a node in the \papermatrix, converging at a field at the bottom of the \papermatrix, and finally merging into \patenttreemap, as if the knowledge is flowing from the broad scientific landscape to the technology fields (Fig.~\ref{fig:visual-design}). 
The width of the flow represents the number of patent citations. The thicker the flow, the heavier the patent category relies on the knowledge from the collection of the connected papers. 

We tend to lay out paper fields with similar patent citations near each other to conveniently explore interdisciplinary patent citations. At the same time, we also keep paper fields with more patent citations near the center of the \papermatrix for a balanced visual appearance with thick flows in the center. To reduce the visual clutter caused by a large number of citation links, we reorder the research fields and route the flows to help reduce line crossing. Formally, the layout procedures described above can be formulated as an optimization problem with the following objective:
\[
      \alpha \sum_{i<j}^{} w_{ij} \left\| x_{Q_{i}} - x_{Q_{j}}\right\|^{2} 
      + \beta \sum_{i=1}^{m} \left\| x_{Q_{i}} - x_{Q^{\prime}_{i}} \right\|^{2} 
      + \gamma \sum_{i=1}^{m}\sum_{j=1}^{n}\left\|x_{Q_{i}} - x_{P_{j}}\right\|^{2} 
\]
where $x$ represents the horizontal position of the paper or patent fields. $\mathbb{Q}$ represents the set of total $m$ paper fields: ${\{ Q_{1}, ..., Q_{m}\}}$. 
$\mathbb{Q^{\prime}}$ denotes the optimally ordered list of $m$ paper fields, sorted by the number of patent citations in the field, which positions fields with the most citations in the center: ${\{ Q^{\prime}_{1}, ..., Q^{\prime}_{m}\}}$. 
$\mathbb{P}$ represents the set of total $n$ patent categories: ${\{ P_{1}, ..., P_{n}\}}$. $w_{ij}$ is the cosine similarity of paper pairs based on patent citation similarity. Intuitively, the first term in the objective function puts paper fields with similar patent citations near each other. 
The second term balances the flows by centralizing paper fields with more patent citations. The third term minimizes the paper-patent citation flow crossings. 
We balance the three parts based on the parameters $\alpha$, $\beta$, and $\gamma$. The flows are rendered using cubic Bézier curves, which are bundled to reduce visual clutter and emphasize important flows. 


\subsection{\secondround{Context Views}}
\label{sec:05_VisualDesign_OtherView}
The system also provides a number of coordinated views for users to explore the connection between science and technology and identify untapped potential more systematically with context information. 

\textbf{\researcheroverviewview.} This view summarizes all researchers in a scatter plot (\secondround{\textbf{\tasktypeoneshort1}, \textbf{\tasktypethreeshort1}}, Fig.~\ref{fig:system-ui}(A)) \secondround{to help experts locate researchers of interest based on research profiles}. Each circle represents a researcher, with the opacity encoding the P-index. We use a contour map to show the distributions of researchers. The x-axis and y-axis indicate two researcher metrics (introduced in \secondround{Section~\ref{sec:04_DataAnalysis}}), which can be interactively changed based on users' preferences. 

\textbf{\researcherstatisticsview}. This view summarizes characteristics and detailed researcher information for the selected group (\secondround{\textbf{\tasktypeoneshort2}}, \textbf{\tasktypethreeshort1}). It supports two visual modes: (1) bar charts and histograms that summarize the demographic information and SciSci metric distribution (Fig.~\ref{fig:system-ui}(B1)); and (2) researcher cards that show the list of researchers (Fig.~\ref{fig:system-ui}(B2)), which can be ranked by different SciSci metrics.  

\textbf{\fieldtimeline}. \secondround{The time dimension is also essential to help users identify trending topics in the dual frontiers.} Thus, we designed two groups of horizon graphs to reveal the temporal evolution of different fields in science and technology (\secondround{\textbf{\tasktypetwoshort2}}, Fig.~\ref{fig:system-ui}(C1)). Each paper or patent field is represented by a horizon graph which shows the temporal evolution in a space-saving way. The x-axis is the timeline. The saturation of the area encodes the papers or patents published or granted in each field every year. Darker color indicates a higher value.

\textbf{\technologyinspectionview}. This view shows additional context information about patent categories (\secondround{\textbf{\tasktypetwoshort3}}, Fig.~\ref{fig:system-ui}(D)). 
\secondround{
In addition to patent categories displayed in the \interplayview, the experts wanted more details on the assignee distribution and patent topics to aid in decision-making (\eg, identifying potential partners and directions for invention commercialization). 
Thus, this view includes a two-level sunburst showing the proportion of assignees and a word cloud in the center showing the keywords of the selected patents.
} 

\textbf{\scienceinspectionview}. \secondround{To facilitate paper-level exploration}, the \scienceinspectionview shows a paper list (\secondround{\textbf{\tasktypetwoshort3}}, Fig.~\ref{fig:system-ui}(E)) that can be ranked based on the statistical matrices introduced in Section 4.2 as well as a list of histograms showing distributions of paper-level metrics.

\subsection{\secondround{{Iterative Design Process}}} 
\label{sec:05_VisualDesign_AlternativeDesign} 
\secondround{
We went through multiple iterations for the design of the major visual components with our experts. 
%
Specifically, for the \interplayview, 
we initially used a graph where each node represented a paper or patent. 
However, it was not scalable for data with large volumes, and experts' feedback indicated that a field-level display was more critical. 
We thus aggregated the graph to the field level as a bipartite graph. 
One further feedback was to unfold papers in the same field into different groups based on patent citations, as they wanted to compare the differences among these paper groups. 
We thus unfolded paper-field nodes into a paper matrix with statistics summarized in each matrix node. 
Finally, our experts asked us to depict the temporal evolution of the dual frontiers for the purpose of identifying trending topics. 
We thus presented three choices: a timeline emphasizing temporal evolution~\cite{burch2011parallel}, and two others emphasizing the citation structure (node-link (Section~\ref{sec:05_VisualDesign_Innovation_View_Interplay_Graph}) and matrix~\cite{burch2013matrix}). 
The experts ultimately prioritized citation structure with the more intuitive node-link representation, leading us to depict the timeline as a secondary data dimension via the horizon graphs. 
}

\section{Evaluation}
\label{sec:06_Evaluation} 
We evaluated \systemname through a quantitative study of the prediction model, two case studies, and interviews with experts. 

\subsection{Quantitative Evaluation of the Prediction Model} 
Due to the class imbalance between papers that receive patent citations and those that do not, we evaluate the performance of our model using the AUC (Area Under the Curve) metric. 
Our final model is chosen from the epoch that produces the highest AUC score on the validation set. 
We evaluate the model by presenting its prediction performance and scalability in the two datasets used in two case studies (Section~\ref{sec:06_Evaluation_CaseStudy}). We apply the prediction pipeline above on the top $K$ patent CPC groups $g$ (denoted as $\mathbb{G}$) in the case studies in Section~\ref{sec:06_Evaluation_CaseStudy}. We focus on the top $50$ patent CPC groups because they cover more than 95\% of patents citing our target papers in the case studies. 

\begin{figure}[tb]
  \centering
  \begin{subfigure}[b]{0.45\columnwidth}
  	\centering
  	\includegraphics[width=\textwidth]{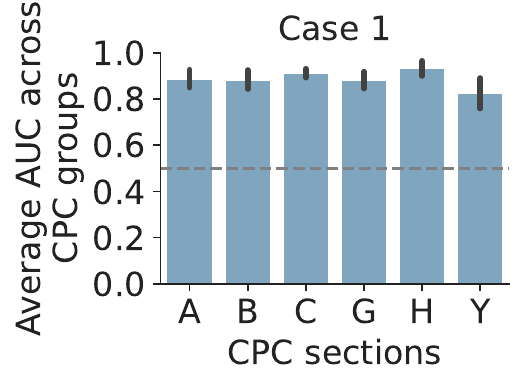}
  	\label{fig:case1_auc}
  \end{subfigure}%
  \hfill%
  \begin{subfigure}[b]{0.45\columnwidth}
  	\centering
  	\includegraphics[width=\textwidth]{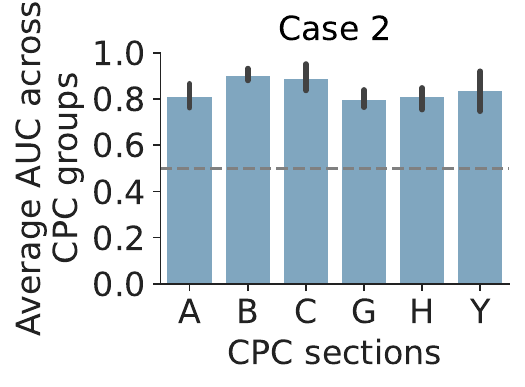}
  	\label{fig:case2_auc}
  \end{subfigure}%
  \vspace{-3mm}
  \subfigsCaption{The prediction performance (AUC) on the test set for two case studies. The average AUC and its 95\% confidence interval are reported for each case and CPC section, based on the performance across CPC groups within that section. The dashed lines at 0.5 indicate the baseline performance of random guesses.
  Overall, the results demonstrate good performance for our prediction tasks.
  }
  \vspace{-0.5cm}\label{fig:case1_case2_auc}
\end{figure}

\begin{itemize}[leftmargin=10pt,topsep=1pt,itemsep=0px]
    \item \textbf{AUC}. We present the AUC results on the test set across CPC groups by CPC section in Fig.~\ref{fig:case1_case2_auc}. The overall prediction performance is good and remains robust across different CPC groups. 
    \item \textbf{Scalability}. The time complexity of GCN has been demonstrated to be linear in the number of graph edges and can scale to millions of edges~\cite{kipf2017semi}. In our two cases, the training time per epoch was less than 10 seconds with CPU-only implementations. We also ran the codes in parallel for different CPC groups to accelerate the prediction process. Moreover, the prediction model is pre-executed and does not impact the visualization system in real time. 
\end{itemize}
    
\subsection{Case Study} 
\label{sec:06_Evaluation_CaseStudy} 
We invited our experts to explore the system. 
\secondround{First, each expert explored the system independently. We then summarized their findings and formed two case studies to demonstrate our system.}

\begin{figure} [tb]
 \centering 
 \includegraphics[width=0.95\linewidth]{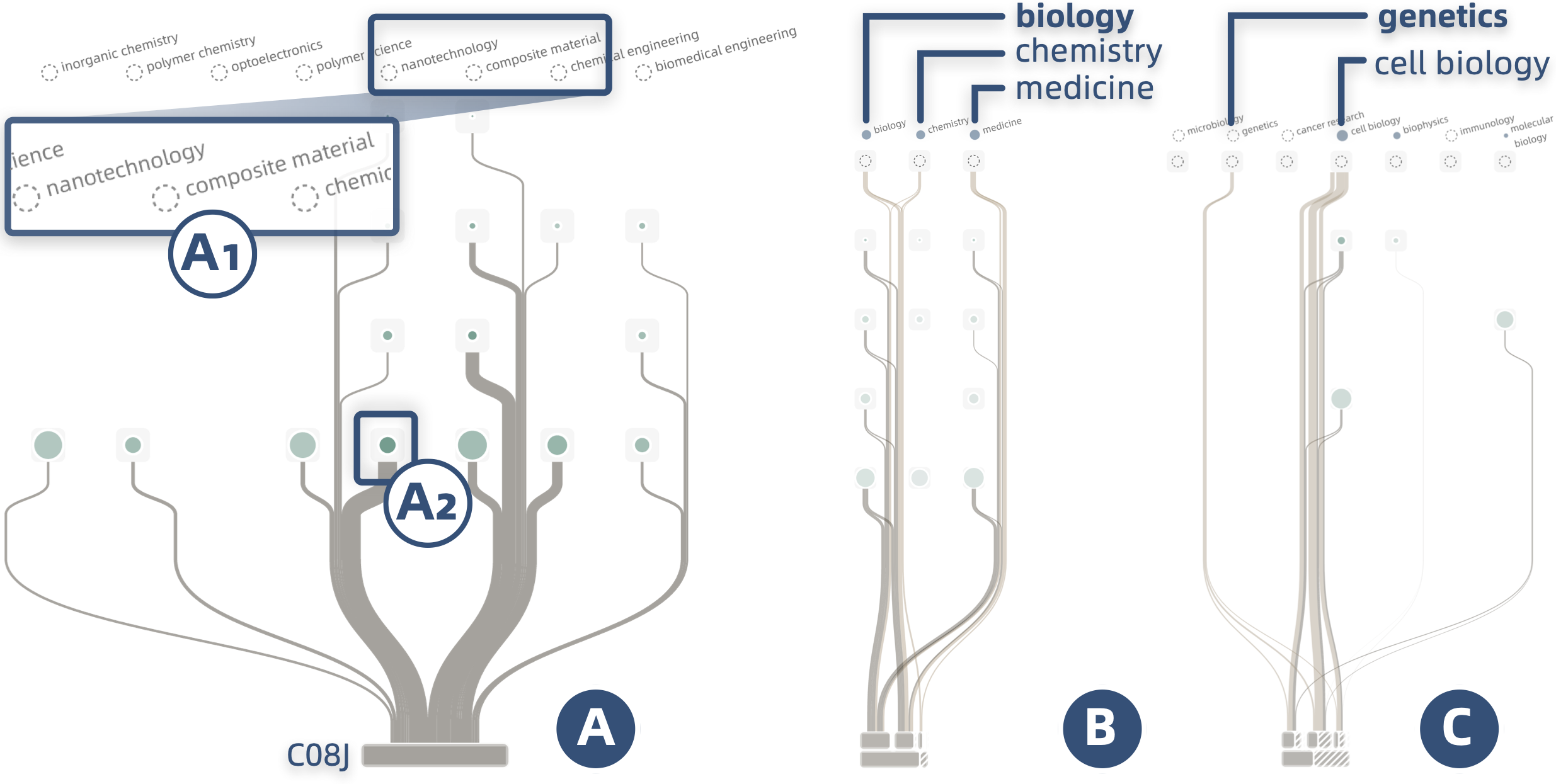}
 \caption{
    The paper-patent citation flow in case 1. 
    (A) \textit{Nanotechnology} and \textit{composite material} are the predominant science fields consumed by patent category \textit{C08J}. 
    (B) The historical flow shows most papers published by
    selected researchers are in basic \textit{biology} journals. 
    (C) The prediction flow shows the papers in \textit{genetics} published by the selected faculty also have a high potential for innovation. 
 } 
 \vspace{-0.6cm}
 \label{fig:case1}
\end{figure}

\begin{figure*} [!htb]
 \centering 
 \vspace{-0.5cm}
 \includegraphics[width=0.99\linewidth]{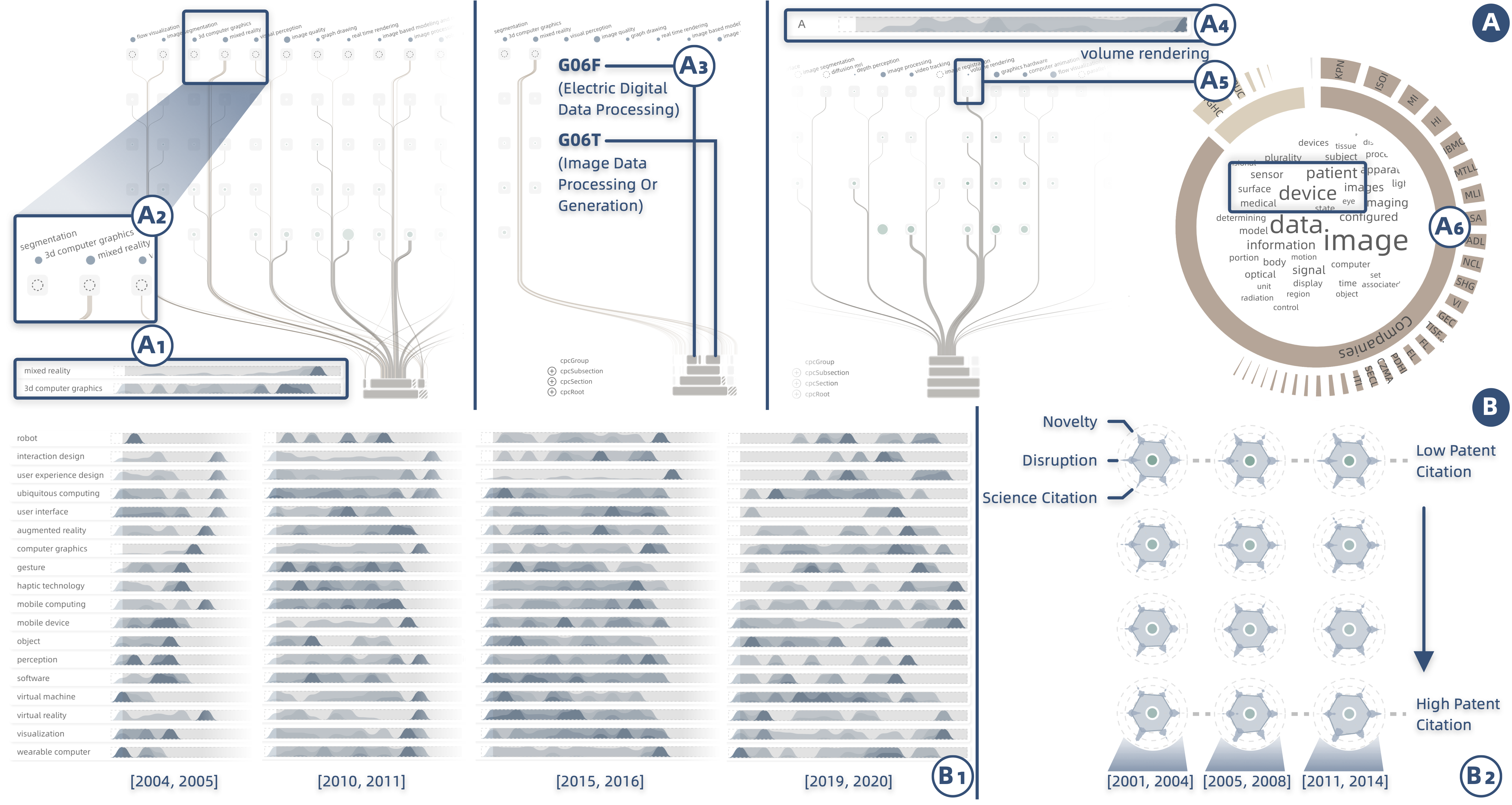}
 \vspace{-0.1cm} 
 \caption{
    Innovation insights in case 2. 
    (A1)-(A2) The spotting hot new frontiers in the VIS community include \textit{mixed reality} and \textit{3d computer graphics}. 
    (A3) \textit{mixed reality} is predicted to be used in a variety of patent categories. 
    (A4)-(A6) Papers in \textit{volume rendering} are still being cited by new patents in medicine. 
    (B1) New technologies are increasingly relying on older scientific knowledge. 
    (B2) Relationships between patent citation and other SciSci metrics: science citations and novelty have positive relationships, while disruption has a negative relationship with patent citations. 
    } 
 \vspace{-0.5cm} 
 \label{fig:case2}
\end{figure*}

\subsubsection{New Opportunities for University Innovation} 
\label{sec:06_Evaluation_CaseStudy_Case1_NU} 
We piloted the system with a premier research university and focused on $461$ faculty who have at least one paper that has been cited by a patent (37K papers in total). 
This case study demonstrates how \systemname enables our experts (\ea and \eb) to uncover new innovation opportunities and facilitate knowledge transfer for researchers. 

\textbf{Overview of the innovation landscape}. 
After loading the data from \universityname, the experts started from the \citationflow (\textbf{\tasktypetwoshort1}, Fig.~\ref{fig:system-ui}(C3)), finding 
\textit{material science}, \textit{biology}, and \textit{chemistry}, as the three predominant disciplines whose knowledge has spurred inventions across many patent categories. 
Highlighting one category \textit{C} (\textit{``Chemistry; Metallurgy''}, Fig.~\ref{fig:system-ui}(C4)), the experts noticed that this category draws from papers in not just \textit{chemistry} but also \textit{material science} and \textit{biology}, emphasizing its interdisciplinary orientation. 
There was also high diversity in the paper-patent citation for \textit{material science} (Fig.~\ref{fig:system-ui}(C5)). 
Zooming in, the experts found at least six major patent categories, with \textit{C08J} (\textit{General Processes of Compounding}) as the most rapidly emerging technology area.
This area primarily consumed knowledge from  \textit{nanotechnology} and \textit{composite material} (Fig.~\ref{fig:case1}(A1)), which indicates that these research topics were important in this university. 
This patent category also relied heavily on papers with large numbers of patent citations (Fig.~\ref{fig:case1}(A2)) in \textit{nanotechnology}, showing their importance in technology development. 
Choosing this cell (\textbf{\tasktypetwoshort3}), the experts found that the top-citation papers were about \textit{polymer nanocomposites}. 
When the experts went to the \technologyinspectionview (\textbf{\tasktypetwoshort3}, Fig.~\ref{fig:system-ui}(D)), they found that unexpectedly, the largest university assignee was KFUPM
(a university in Saudi Arabia, Fig.~\ref{fig:system-ui}(D1)) rather than \universityname itself, and that 
this university just started to cite papers in \universityname in recent years. 
This finding represented fresh insights for the experts: \textit{``in most cases, we expect our own university to be the largest assignee citing our papers. 
So this finding is rather unexpected. Maybe there are collaboration opportunities with KFUPM, especially in \textit{nanotechnology}.''} 

\textbf{Uncover hidden talents}. 
The experts are also interested in uncovering hidden talents and untapped innovation potential. 
Indeed, when examining the \researcheroverviewview (\textbf{\tasktypeoneshort1}, Fig.~\ref{fig:system-ui}(A)), plotting the number of invention disclosures vs. patent-cited papers for each individual, our experts immediately discovered a fascinating insight: 
at the bottom of Fig.~\ref{fig:system-ui}(A1) lay an interesting group of researchers. 
They themselves had no invention disclosures, yet their papers had been cited frequently by other patents. 
\secondround{\textit{``Who are these people?''}} our experts asked immediately. 
Zooming in on the \researcherstatisticsview (\textbf{\tasktypeoneshort2}, Fig.~\ref{fig:system-ui}(B1)) revealed that most of them were full professors at \secondround{the} medical school. 
Most of their papers were published in basic biology journals (Fig.~\ref{fig:case1}(B)), yet surprisingly, they were being cited heavily by companies, finding widespread uses in the private sector (Fig.~\ref{fig:system-ui}(D2)). 


The experts thus ranked these researchers by P-index to locate those with high commercialization potential. 
They quickly noticed the second faculty (Fig.~\ref{fig:system-ui}(B2)) who had a rather high P-index and was the only female faculty among the top P-index researchers. 
She had no inventions, but many of her papers were cited by patents from private companies. 
From the \interplayview (\textbf{\tasktypetwoshort1}, Fig.~\ref{fig:case1}(C)), the experts gathered that most of her papers were in \textit{cell biology} but were increasingly cited by 
patents in the \textit{C12N} 
(\textit{Microorganisms or Enzymes; Mutation or Genetic Engineering}) 
category (\textbf{\tasktypetwoshort2}, Fig.~\ref{fig:system-ui}(C6)). And many of these patents were from the same company, namely \textit{Immatics Biotechnologies}
in Germany (\textbf{\tasktypetwoshort3}, Fig.~\ref{fig:system-ui}(D3)). 
This discovery prompted our experts to hold an immediate follow-up conversation with the faculty member. 
It turned out that while this faculty had done a sabbatical in Germany, she was completely unaware of this company, or the fact that they were drawing heavily on her research.
Using our system, the experts further showed the faculty member other prediction results on which of her other papers in \textit{genetics} (\textbf{\tasktypethreeshort1}, Fig.~\ref{fig:case1}(C)) revealed a high potential for innovation. 
Two weeks following the conversation, she submitted new invention disclosures to the university, for the first time in her career! 

The case shows that \systemname is highly effective in uncovering new innovation potential and opportunities in research institutions.

\subsubsection{Innovation Insights in VIS Communities}
\label{sec:06_Evaluation_CaseStudy_Case2_VIS} 
Our experts (\ec and \ed) used \textit{visualization} as an exemplary field to explore the dual frontiers of science and technology. 
Specifically, we identified 2016 top-publishing researchers based on 6 major journals and 13 leading conferences in VIS and analyzed all their publications. 

\noindent \textbf{Spotting hot new frontiers}. 
The experts first focused on the topic of \textit{visualization}. 
In the paper field timeline (\textbf{\tasktypetwoshort2}, Fig.~\ref{fig:case2}(A1)), two topics quickly stood out: \textit{mixed reality} and \textit{3d computer graphics}. 
In addition to being highly popular, these two topics were paired together in the \interplayview (Fig.~\ref{fig:case2}(A2)),  suggesting that they were frequently co-cited by similar patent categories. 
The experts then filtered recent papers with high prediction scores (\textbf{\tasktypethreeshort1}). 
Interestingly, while other areas (\eg, \textit{image processing} and \textit{real time rendering}) had tended to dominate the field, recent papers in \textit{mixed reality} are characterized by some of the highest prediction scores (Fig.~\ref{fig:case2}(A2)), with prediction flows coming from a variety of patenting domains (\eg, \textit{G06F} (\textit{Electric Digital Data Processing}) and \textit{G06T} (\textit{Image Data Processing})) (Fig.~\ref{fig:case2}(A3)). 
\textit{"This really speaks to the application potential of 'mixed reality,'"} as our experts commented.  
%
Our experts further noticed that the patent category \textit{A} (\textit{Human Necessities}) was rising in popularity (Fig.~\ref{fig:case2}(A4)). 
Zooming in, they found that \textit{volume rendering} was the most applied science topic (\textbf{\tasktypetwoshort1}, Fig.~\ref{fig:case2}(A5)) and was primarily used in \textit{A61B} (\textit{Diagnosis; Surgery}). 
Highlighting \textit{A61B} and in the \technologyinspectionview (Fig.~\ref{fig:case2}(A6)), the experts found that these patents were related to medical devices. 
Curious about what papers were highly cited by patents, they clicked the paper cell in the last row (\textbf{\tasktypetwoshort3}). 
Somewhat unexpectedly, the highly cited papers in this emerging patent category were not new papers; rather, they were canonical papers in the field (\eg, ~\cite{fattal2001variational}). 
\textit{``Interesting! Even after ten years, papers in `volume rendering' are still being cited by new patents. These papers are canons in the field. They have a lasting impact on technology development.''} 

\noindent \textbf{New vs. canonical knowledge?} 
Intrigued by the preceding findings, the experts returned their attention to \textit{human computer interaction} and filtered patent application years based on three ranges in the early, mid, and end of the period between 2001 and 2020 (\textbf{\tasktypetwoshort2}). 
In recent years (\eg, [2019, 2020] and [2015, 2016], Fig.~\ref{fig:case2}(B1)), the patent citation of old and new papers was diverse. 
Patents either cited new paper fields or cited fallen fields many years ago. 
However, when time went back ten more years ago, patents tended to cite more new paper fields at that time (\eg, [2010, 2011] and [2004, 2005], Fig.~\ref{fig:case2}(B1)). 
\textit{``This finding supports our hypothesis that new technologies are increasingly relying on older scientific knowledge.''}

\noindent \textbf{What kinds of papers tend to see greater uses in technology?} 
The experts also explored the characteristics of papers that are heavily cited by patents (\textbf{\tasktypetwoshort1}). 
To obtain a general view, the experts zoomed out and focused on one of the most prominent fields, \textit{computer vision}, as an example. 
%
Inspecting papers published in different periods (Fig.~\ref{fig:case2}(B2)), they found that 
generally, science citations and novelty had positive relationships with patent citations, whereas the disruption score was negatively correlated with patent citations.  
\textit{``These results are consistent with our recent findings~\cite{yin2022public}. 
Papers that are highly valued within science also see greater practical uses. 
At the same time, these results also raise new research questions regarding the relationships between novelty, disruption, and patent citations.''}

Overall, these cases demonstrate the effectiveness of our system in navigating multiple dimensions of data and uncovering new insights. 

\subsection{Expert Interview}
\label{sec:06_Evaluation_ExpertInterview} 
We collected feedback from experts in Section~\ref{sec:03_Background_ProblemFormulation} and interviewed six external experts who had used \systemname for the first time. 
\newea and \neweb are innovation managers in the TTO of a university. 
\newec and \newed are researchers in SciSci. 
Although the target users of \systemname are the above two groups, the dataset in case 2 may also interest VIS researchers. 
Thus, we also included two VIS researchers (\newee and \newef) who have three-year visualization expertise. 
%
Each interview with an external expert lasted about 90 minutes. 
We first briefly introduced the project background, including the analytical tasks and data sources. 
Then we used case 1 in Section~\ref{sec:06_Evaluation_CaseStudy_Case1_NU} to demonstrate the system workflow and visual encodings. 
Third, they were asked to explore the system in a think-aloud manner. 
Finally, we had a semi-structured interview. 
We took notes on their comments and findings during the process. 
\secondround{The feedback from the two groups of experts is summarized below.} 

\textbf{System Workflow.} 
All experts appreciated the clear workflow. 
They were able to learn the system logic quickly, from researcher identification to paper-patent citation exploration. 
We also observed diverse analysis focuses between experts from different fields. 
Innovation managers from the TTO focused more on researchers and their research fields, 
while SciSci and VIS experts tended to start directly with scientific fields. 
As \newec said, \textit{``we are more interested in the general findings over scientific fields.''}
Nevertheless, they agreed that the current workflow was able to meet different needs through filtering schemes. 

\textbf{SciSci Metrics and Prediction Model.} 
All of the experts showed interest in these metrics. 
Those from the TTO were particularly interested in the \textit{P-index} obtained from the prediction model. 
\neweb was excited about this straightforward approach to locating faculty with high commercialization potential. 
\ea suggested making the prediction more transparent to help them understand the mechanism behind the prediction. 
\newee and \newef found metrics such as \textit{team size} and \textit{novelty} interesting, \textit{``these metrics provide us new perspectives to look at papers besides paper citations.''} 

\textbf{Visualization and Interactions.} 
The experts noted that the visual components in the system were intuitive and satisfied all the analytical tasks. 
Many of them appreciated the \interplayview. 
\ea and \newef reported that it took some time to understand the encoding, but it eventually became very useful and intuitive. 
\newec especially appreciated the y-axis in the paper matrix, as she could locate papers with high patent citations more quickly. 
\neweb and \newee liked the \technologyinspectionview and also suggested, \secondround{\textit{``it would be more interesting if we could check the relationships between these assignees.''}} 
\newec and \newef liked the glyph design, \secondround{\textit{``it makes the comparison between paper groups much easier. 
But it would be great if labels could be added to show the meaning of dimensions.''
}}
\ea also mentioned the timeline as important to check recent hot areas as a temporal indicator of innovation potential. 

\textbf{Suggestions.} 
Despite universities and research fields, \ed also wanted to filter researchers at the regional level to compare innovation. 
\ea suggested using text analysis to reveal more detail about paper-patent citations, such as the distinction between strong citations (\ie, cite the core knowledge in the paper) and weak citations (\ie, cite a paper in the background). 
\section{Discussion}
\label{sec:07_Discussion} 
This section discusses the significance and generalizability, lessons learned, and limitations of our work. 

\textbf{Significance and Generalizability.} 
Science provides a foundation for many practical applications in human society, but the pathway through which basic understanding leads to technological development is neither visible nor intuitive. 
Consider Einstein's theory of general relativity, deemed as the discovery of the 20th century. Among the myriad innovations it spurred, it proved essential for the Global Positioning System \secondround{(GPS)} through time dilation corrections. The GPS system then provided the technical foundation for applications such as Uber. 
The ability to effectively trace and visualize the evolving dual frontiers of science and technology is therefore crucial to understanding how science drives practical applications and leads to rising standards of living. 
Our system not only fulfills our original design purposes, allowing users to better identify the sources of technical inventions and understand the holistic impact of scientific research; 
it also enables an array of new applications for researchers and research institutions, ranging from identifying untapped innovation potentials within an institution to forging new partnership opportunities between science and industry. 
Moreover, the proposed SciSci metrics, prediction model, and visualization system can be adapted for studying other upstream (\eg, funding) and downstream (\eg, policy documents) linkages to science~\cite{yin2022public}. 
When adapting the system to other domains, we suggest replacing the patent data with other upstream or downstream data and refining the metrics and prediction model to suit specific scenarios. 

\textbf{Lessons Learned.} 
The design study with experts from multiple fields provides valuable insights into conducting interdisciplinary research. 
First, regularly discussing data insights with experts substantially accelerates progress through timely clarifications of analytical goals. 
We started with exploratory analysis and used static charts to discuss initial findings. 
This practice helped us quickly verify insights, facilitating adjustments to our analytical goals. 
Moreover, the initial findings also inform experts about their practice.
During the process, our experts identified researchers with high commercialization potential (Section~\ref{sec:06_Evaluation_CaseStudy_Case1_NU}) and helped them to submit invention disclosures. 
%
Second, the visual design of data with a complex structure requires involving multiple design choices. 
\secondround{
Experts often lack clarity on the level of detail the data should be presented, generally desiring maximum information. 
It is helpful to offer alternatives at varied granularities and prioritize data dimensions based on their importance for analysis goals. 
}

\textbf{Limitations.} 
Our system is not without limitations. 
First, the current prediction model relies on citations between papers without considering those between papers and patents. 
Future work may use other GNN models designed for heterogeneous graphs, which can incorporate different types of nodes (\eg, papers and patents). 
\secondround{
In addition, the P-index, derived from a GNN model trained on papers published between 2001 and 2014, can be improved by expanding the training dataset with recent papers. We also plan to keep collaborating with TTO experts for model validation and enhancement. 
Second, due to restrictions on data availability, the scope of our study is limited to a single university or specific field with patent data from the USPTO. Future research may find patent data worldwide and extend the partnership to other universities in order to encompass a wider innovation landscape. 
Third, 
the current system exhibits some latency when computing results in real time, particularly with large datasets. 
We intend to resolve this issue using progressive visual analytics~\cite{angelini2018review, fekete2019progressive}. 
%
Lastly, 
to ensure the system's long-term utility, we plan to regularly maintain the system and the model with the most recent data (\eg, OpenAlex~\cite{openalex}).}

\section{Conclusion and Future Directions}
\label{sec:08_Conclusion} 
This paper presents \systemname, a first-of-its-kind visualization system for researchers and research institutions to explore the complex interactions between science and technology. 
It supports analyzing multiple entities (\eg, researchers, papers, and patents) through descriptive and predictive analyses. 
Coordinated views with intuitive interactions are developed to support analysis. 
Two case studies, expert interviews, and our engagement project with a partner university demonstrate the substantial utility and potential impact of our system. 
\secondround{In the future, we plan to integrate more data and release an online system for public use.} 

This work opens up several fruitful future directions, especially at the intersection of data visualization and the science of science. For example, beyond understanding the impact of science on technological development, visualization approaches would prove fruitful for analyzing the broad uses of science across several crucial downstream applications in society, tracing how science is used in the hallways of governments through science-policy linkages, as well as how science enables life-saving drugs and therapeutics by incorporating clinical trials data. Further, data visualization techniques can also help us better understand the multi-dimensional impacts of funding on scientific progress and individual careers. 
Developing novel and efficient visual analytics systems to analyze large-scale data, spanning from upstream funding to science to downstream applications, will usefully serve a diverse range of stakeholders, including university leaders, private companies and investors, funding agencies, policymakers, and researchers themselves.  
Given the crucial role of science in improving the human condition, such systems have the potential to unlock enormous value for science---and for society at large. 




\balance
\end{spacing}



\maketitle

\newpage
\acknowledgments{
  \cameraready{ 
  The authors wish to thank Benjamin F. Jones for his helpful comments. They also would like to express their appreciation to anonymous reviewers for their valuable comments. 
  This work is supported by the Air Force Office of Scientific Research under award numbers FA9550-17-1-0089 and FA9550-19-1-0354, the Alfred P. Sloan Foundation G-2019-12485, and the Future Wanxiang Foundation. 
  }
} 

\suppmaterials{
  The authors provide the following materials at \url{https://kellogg-cssi.github.io/InnovationInsights/}: 
  (1) a video introducing the research background and system interface, 
  (2) a video presenting a case study of the system, 
  and (3) a document describing two datasets used in case studies. 
}

\bibliographystyle{abbrv-doi-hyperref-narrow}
\bibliography{main}

\begin{thebibliography}{10}
\renewcommand*{\sfdefault}{PTSansNarrow-TLF}

\bibitem{cpcscheme}
{Cooperative Patent Classification (CPC)}.
\newblock
  \url{https://www.uspto.gov/patents/search/classification-standards-and-development}.

\bibitem{openalex}
{OpenAlex}.
\newblock \url{https://openalex.org/}.

\bibitem{patentsview}
{PatentsView}.
\newblock \url{https://patentsview.org/}.

\bibitem{uspto}
{USPTO}.
\newblock \url{https://www.uspto.gov/}.

\bibitem{abdelaal2022comparative}
M.~Abdelaal, N.~D. Schiele, K.~Angerbauer, K.~Kurzhals, M.~Sedlmair, and
  D.~Weiskopf.
\newblock Comparative evaluation of bipartite, node-link, and matrix-based
  network representations.
\newblock {\em IEEE Transactions on Visualization and Computer Graphics},
  29(1):896--906, 2022.
  \href{https://doi.org/10.1109/10.1109/TVCG.2022.3209427}
{doi: \textsf{%
10\hspace{.1pt}\discretionary{.}{%
}{.}\hspace{.4pt}1109\discretionary{/}{%
}{/}10\hspace{.1pt}\discretionary{.}{%
}{.}\hspace{.4pt}1109\discretionary{/}{%
}{/}TVCG\hspace{.1pt}\discretionary{.}{%
}{.}\hspace{.4pt}2022\hspace{.1pt}\discretionary{.}{%
}{.}\hspace{.4pt}3209427}}


\bibitem{ahmadpoor2017dual}
M.~Ahmadpoor and B.~F. Jones.
\newblock The dual frontier: Patented inventions and prior scientific advance.
\newblock {\em Science}, 357(6351):583--587, 2017.
  \href{https://doi.org/10.1126/science.aam9527}
{doi: \textsf{%
10\hspace{.1pt}\discretionary{.}{%
}{.}\hspace{.4pt}1126\discretionary{/}{%
}{/}science\hspace{.1pt}\discretionary{.}{%
}{.}\hspace{.4pt}aam9527}}


\bibitem{angelini2018review}
M.~Angelini, G.~Santucci, H.~Schumann, and H.-J. Schulz.
\newblock A review and characterization of progressive visual analytics.
\newblock In {\em Informatics}, vol.~5, p.~31, 2018.
  \href{https://doi.org/10.3390/informatics5030031}
{doi: \textsf{%
10\hspace{.1pt}\discretionary{.}{%
}{.}\hspace{.4pt}3390\discretionary{/}{%
}{/}informatics5030031}}


\bibitem{ankam2012exploring}
E.~Ankam, W.~Dou, D.~Strumsky, D.~X. Wang, T.~Rabinowitz, and W.~Zadrozny.
\newblock Exploring emerging technologies using patent data and patent
  classifications.
\newblock In {\em Proceedings of the CHI Conference on Human Factors in
  Computing Systems}, 2012.

\bibitem{beck2017taxonomy}
F.~Beck, M.~Burch, S.~Diehl, and D.~Weiskopf.
\newblock A taxonomy and survey of dynamic graph visualization.
\newblock In {\em Computer Graphics Forum}, vol.~36, pp. 133--159. Wiley Online
  Library, 2017. \href{https://doi.org/10.1111/cgf.12791}
{doi: \textsf{%
10\hspace{.1pt}\discretionary{.}{%
}{.}\hspace{.4pt}1111\discretionary{/}{%
}{/}cgf\hspace{.1pt}\discretionary{.}{%
}{.}\hspace{.4pt}12791}}


\bibitem{borner2010atlas}
K.~B{\"o}rner.
\newblock {\em Atlas of science: Visualizing what we know}.
\newblock Mit Press, 2010. \href{https://doi.org/10.1007/s11192-011-0409-7}
{doi: \textsf{%
10\hspace{.1pt}\discretionary{.}{%
}{.}\hspace{.4pt}1007\discretionary{/}{%
}{/}s11192\discretionary{%
}{-}{-}011\discretionary{%
}{-}{-}0409\discretionary{%
}{-}{-}7}}


\bibitem{boyack2000analysis}
K.~W. Boyack, B.~N. Wylie, G.~S. Davidson, and D.~K. Johnson.
\newblock Analysis of patent databases using {VxInsight}.
\newblock Technical report, \secondround{Sandia National Lab. (SNL-NM),
  Albuquerque, NM (United States); Sandia National Lab. (SNL-CA), Livermore, CA
  (United States)}, 2000.

\bibitem{buchin2011flow}
K.~Buchin, B.~Speckmann, and K.~Verbeek.
\newblock Flow map layout via spiral trees.
\newblock {\em \secondround{IEEE Transactions on Visualization and Computer
  Graphics}}, 17(12):2536--2544, 2011.
  \href{https://doi.org/10.1109/TVCG.2011.202}
{doi: \textsf{%
10\hspace{.1pt}\discretionary{.}{%
}{.}\hspace{.4pt}1109\discretionary{/}{%
}{/}TVCG\hspace{.1pt}\discretionary{.}{%
}{.}\hspace{.4pt}2011\hspace{.1pt}\discretionary{.}{%
}{.}\hspace{.4pt}202}}


\bibitem{burch2013matrix}
M.~Burch, B.~Schmidt, and D.~Weiskopf.
\newblock A matrix-based visualization for exploring dynamic compound digraphs.
\newblock In {\em 17th International Conference on Information Visualisation},
  pp. 66--73. IEEE, 2013. \href{https://doi.org/10.1109/IV.2013.8}
{doi: \textsf{%
10\hspace{.1pt}\discretionary{.}{%
}{.}\hspace{.4pt}1109\discretionary{/}{%
}{/}IV\hspace{.1pt}\discretionary{.}{%
}{.}\hspace{.4pt}2013\hspace{.1pt}\discretionary{.}{%
}{.}\hspace{.4pt}8}}


\bibitem{burch2011parallel}
M.~Burch, C.~Vehlow, F.~Beck, S.~Diehl, and D.~Weiskopf.
\newblock Parallel edge splatting for scalable dynamic graph visualization.
\newblock {\em \secondround{IEEE Transactions on Visualization and Computer
  Graphics}}, 17(12):2344--2353, 2011.
  \href{https://doi.org/10.1109/TVCG.2011.226}
{doi: \textsf{%
10\hspace{.1pt}\discretionary{.}{%
}{.}\hspace{.4pt}1109\discretionary{/}{%
}{/}TVCG\hspace{.1pt}\discretionary{.}{%
}{.}\hspace{.4pt}2011\hspace{.1pt}\discretionary{.}{%
}{.}\hspace{.4pt}226}}


\bibitem{bush1990science}
V.~Bush.
\newblock {\em Science--the {Endless Frontier}: a Report to the President on a
  Program for Postwar Scientific Research}, vol.~90.
\newblock National Science Foundation, 1990.

\bibitem{cao2023breaking}
H.~Cao, Y.~Lu, Y.~Deng, D.~A. McFarland, and M.~S. Bernstein.
\newblock Breaking out of the ivory tower: A large-scale analysis of patent
  citations to {HCI} research.
\newblock In {\em Proceedings of the CHI Conference on Human Factors in
  Computing Systems}, 2023. \href{https://doi.org/10.1145/3544548.3581108}
{doi: \textsf{%
10\hspace{.1pt}\discretionary{.}{%
}{.}\hspace{.4pt}1145\discretionary{/}{%
}{/}3544548\hspace{.1pt}\discretionary{.}{%
}{.}\hspace{.4pt}3581108}}


\bibitem{chan2018v}
G.~Y.-Y. Chan, P.~Xu, Z.~Dai, and L.~Ren.
\newblock Vibr: Visualizing bipartite relations at scale with the minimum
  description length principle.
\newblock {\em \secondround{IEEE Transactions on Visualization and Computer
  Graphics}}, 25(1):321--330, 2018.
  \href{https://doi.org/10.1109/TVCG.2018.2864826}
{doi: \textsf{%
10\hspace{.1pt}\discretionary{.}{%
}{.}\hspace{.4pt}1109\discretionary{/}{%
}{/}TVCG\hspace{.1pt}\discretionary{.}{%
}{.}\hspace{.4pt}2018\hspace{.1pt}\discretionary{.}{%
}{.}\hspace{.4pt}2864826}}


\bibitem{chen2003mapping}
C.~Chen.
\newblock {\em Mapping scientific frontiers: The quest for knowledge
  visualization}.
\newblock Springer, 2003. \href{https://doi.org/10.1007/978-1-4471-5128-9}
{doi: \textsf{%
10\hspace{.1pt}\discretionary{.}{%
}{.}\hspace{.4pt}1007\discretionary{/}{%
}{/}978\discretionary{%
}{-}{-}1\discretionary{%
}{-}{-}4471\discretionary{%
}{-}{-}5128\discretionary{%
}{-}{-}9}}


\bibitem{chen2021vis30k}
J.~Chen, M.~Ling, R.~Li, P.~Isenberg, T.~Isenberg, M.~Sedlmair, T.~M{\"o}ller,
  R.~S. Laramee, H.-W. Shen, K.~W{\"u}nsche, et~al.
\newblock {VIS30K}: A collection of figures and tables from {IEEE}
  {Visualization} conference publications.
\newblock {\em \secondround{IEEE Transactions on Visualization and Computer
  Graphics}}, 27(9):3826--3833, 2021.
  \href{https://doi.org/10.1109/TVCG.2021.3054916}
{doi: \textsf{%
10\hspace{.1pt}\discretionary{.}{%
}{.}\hspace{.4pt}1109\discretionary{/}{%
}{/}TVCG\hspace{.1pt}\discretionary{.}{%
}{.}\hspace{.4pt}2021\hspace{.1pt}\discretionary{.}{%
}{.}\hspace{.4pt}3054916}}


\bibitem{specter_cohan_2020}
A.~Cohan, S.~Feldman, I.~Beltagy, D.~Downey, and D.~Weld.
\newblock {SPECTER}: Document-level representation learning using
  citation-informed transformers.
\newblock In {\em ACL}, 2020. \href{https://doi.org/10.48550/arXiv.2004.07180}
{doi: \textsf{%
10\hspace{.1pt}\discretionary{.}{%
}{.}\hspace{.4pt}48550\discretionary{/}{%
}{/}arXiv\hspace{.1pt}\discretionary{.}{%
}{.}\hspace{.4pt}2004\hspace{.1pt}\discretionary{.}{%
}{.}\hspace{.4pt}07180}}


\bibitem{dattolo2022authoring}
A.~Dattolo, M.~Corbatto, and M.~Angelini.
\newblock Authoring and reviewing bibliographies: Design and development of a
  visual analytics online platform.
\newblock {\em IEEE Access}, 10:21631--21645, 2022.
  \href{https://doi.org/10.1109/ACCESS.2022.3153027}
{doi: \textsf{%
10\hspace{.1pt}\discretionary{.}{%
}{.}\hspace{.4pt}1109\discretionary{/}{%
}{/}ACCESS\hspace{.1pt}\discretionary{.}{%
}{.}\hspace{.4pt}2022\hspace{.1pt}\discretionary{.}{%
}{.}\hspace{.4pt}3153027}}


\bibitem{deng2022visimages}
D.~Deng, Y.~Wu, X.~Shu, J.~Wu, S.~Fu, W.~Cui, and Y.~Wu.
\newblock {VisImages}: A fine-grained expert-annotated visualization dataset.
\newblock {\em IEEE Transactions on Visualization \& Computer Graphics},
  1(01):1--1, 2022. \href{https://doi.org/10.1109/TVCG.2022.3155440}
{doi: \textsf{%
10\hspace{.1pt}\discretionary{.}{%
}{.}\hspace{.4pt}1109\discretionary{/}{%
}{/}TVCG\hspace{.1pt}\discretionary{.}{%
}{.}\hspace{.4pt}2022\hspace{.1pt}\discretionary{.}{%
}{.}\hspace{.4pt}3155440}}


\bibitem{deng2022multilevel}
Z.~Deng, S.~Chen, X.~Xie, G.~Sun, M.~Xu, D.~Weng, and Y.~Wu.
\newblock Multilevel visual analysis of aggregate geo-networks.
\newblock {\em IEEE Transactions on Visualization and Computer Graphics}, pp.
  1--16, 2022. \href{https://doi.org/10.1109/TVCG.2022.3229953}
{doi: \textsf{%
10\hspace{.1pt}\discretionary{.}{%
}{.}\hspace{.4pt}1109\discretionary{/}{%
}{/}TVCG\hspace{.1pt}\discretionary{.}{%
}{.}\hspace{.4pt}2022\hspace{.1pt}\discretionary{.}{%
}{.}\hspace{.4pt}3229953}}


\bibitem{dong2019vistory}
A.~Dong, W.~Zeng, X.~Chen, and Z.~Cheng.
\newblock {VIStory}: Interactive storyboard for exploring visual information in
  scientific publications.
\newblock In {\em Proceedings of the 12th International Symposium on Visual
  Information Communication and Interaction}, pp. 1--8, 2019.
  \href{https://doi.org/10.1145/3356422.3356430}
{doi: \textsf{%
10\hspace{.1pt}\discretionary{.}{%
}{.}\hspace{.4pt}1145\discretionary{/}{%
}{/}3356422\hspace{.1pt}\discretionary{.}{%
}{.}\hspace{.4pt}3356430}}


\bibitem{dork2012pivotpaths}
M.~D{\"o}rk, N.~H. Riche, G.~Ramos, and S.~Dumais.
\newblock {PivotPaths}: Strolling through faceted information spaces.
\newblock {\em \secondround{IEEE Transactions on Visualization and Computer
  Graphics}}, 18(12):2709--2718, 2012.
  \href{https://doi.org/10.1109/TVCG.2012.252}
{doi: \textsf{%
10\hspace{.1pt}\discretionary{.}{%
}{.}\hspace{.4pt}1109\discretionary{/}{%
}{/}TVCG\hspace{.1pt}\discretionary{.}{%
}{.}\hspace{.4pt}2012\hspace{.1pt}\discretionary{.}{%
}{.}\hspace{.4pt}252}}


\bibitem{dou2013hierarchicaltopics}
W.~Dou, L.~Yu, X.~Wang, Z.~Ma, and W.~Ribarsky.
\newblock {HierarchicalTopics}: Visually exploring large text collections using
  topic hierarchies.
\newblock {\em \secondround{IEEE Transactions on Visualization and Computer
  Graphics}}, 19(12):2002--2011, 2013.
  \href{https://doi.org/10.1109/TVCG.2013.162}
{doi: \textsf{%
10\hspace{.1pt}\discretionary{.}{%
}{.}\hspace{.4pt}1109\discretionary{/}{%
}{/}TVCG\hspace{.1pt}\discretionary{.}{%
}{.}\hspace{.4pt}2013\hspace{.1pt}\discretionary{.}{%
}{.}\hspace{.4pt}162}}


\bibitem{fattal2001variational}
R.~Fattal and D.~Lischinski.
\newblock Variational classification for visualization of {3D} ultrasound data.
\newblock In {\em Proceedings Visualization. VIS'01.}, pp. 403--410. IEEE,
  2001. \href{https://doi.org/10.1109/VISUAL.2001.964539}
{doi: \textsf{%
10\hspace{.1pt}\discretionary{.}{%
}{.}\hspace{.4pt}1109\discretionary{/}{%
}{/}VISUAL\hspace{.1pt}\discretionary{.}{%
}{.}\hspace{.4pt}2001\hspace{.1pt}\discretionary{.}{%
}{.}\hspace{.4pt}964539}}


\bibitem{federico2016survey}
P.~Federico, F.~Heimerl, S.~Koch, and S.~Miksch.
\newblock A survey on visual approaches for analyzing scientific literature and
  patents.
\newblock {\em \secondround{IEEE Transactions on Visualization and Computer
  Graphics}}, 23(9):2179--2198, 2016.
  \href{https://doi.org/10.1109/TVCG.2016.2610422}
{doi: \textsf{%
10\hspace{.1pt}\discretionary{.}{%
}{.}\hspace{.4pt}1109\discretionary{/}{%
}{/}TVCG\hspace{.1pt}\discretionary{.}{%
}{.}\hspace{.4pt}2016\hspace{.1pt}\discretionary{.}{%
}{.}\hspace{.4pt}2610422}}


\bibitem{fekete2019progressive}
J.-D. Fekete, D.~Fisher, A.~Nandi, and M.~Sedlmair.
\newblock Progressive data analysis and visualization, 2019.
  \href{https://doi.org/10.4230/DagRep.8.10.1}
{doi: \textsf{%
10\hspace{.1pt}\discretionary{.}{%
}{.}\hspace{.4pt}4230\discretionary{/}{%
}{/}DagRep\hspace{.1pt}\discretionary{.}{%
}{.}\hspace{.4pt}8\hspace{.1pt}\discretionary{.}{%
}{.}\hspace{.4pt}10\hspace{.1pt}\discretionary{.}{%
}{.}\hspace{.4pt}1}}


\bibitem{Fey/Lenssen/2019}
M.~Fey and J.~E. Lenssen.
\newblock Fast graph representation learning with {PyTorch Geometric}.
\newblock In {\em ICLR Workshop on Representation Learning on Graphs and
  Manifolds}, 2019. \href{https://doi.org/10.48550/arXiv.1903.02428}
{doi: \textsf{%
10\hspace{.1pt}\discretionary{.}{%
}{.}\hspace{.4pt}48550\discretionary{/}{%
}{/}arXiv\hspace{.1pt}\discretionary{.}{%
}{.}\hspace{.4pt}1903\hspace{.1pt}\discretionary{.}{%
}{.}\hspace{.4pt}02428}}


\bibitem{fortunato2018science}
S.~Fortunato, C.~T. Bergstrom, K.~B{\"o}rner, J.~A. Evans, D.~Helbing,
  S.~Milojevi{\'c}, A.~M. Petersen, F.~Radicchi, R.~Sinatra, B.~Uzzi, et~al.
\newblock Science of science.
\newblock {\em Science}, 359(6379):eaao0185, 2018.
  \href{https://doi.org/10.1126/science.aao0185}
{doi: \textsf{%
10\hspace{.1pt}\discretionary{.}{%
}{.}\hspace{.4pt}1126\discretionary{/}{%
}{/}science\hspace{.1pt}\discretionary{.}{%
}{.}\hspace{.4pt}aao0185}}


\bibitem{gonzalez2023landscape}
R.~Gonz{\'a}lez-M{\'a}rquez, L.~Schmidt, B.~M. Schmidt, P.~Berens, and
  D.~Kobak.
\newblock The landscape of biomedical research.
\newblock {\em bioRxiv}, pp. 2023--04, 2023.
  \href{https://doi.org/10.1101/2023.04.10.536208}
{doi: \textsf{%
10\hspace{.1pt}\discretionary{.}{%
}{.}\hspace{.4pt}1101\discretionary{/}{%
}{/}2023\hspace{.1pt}\discretionary{.}{%
}{.}\hspace{.4pt}04\hspace{.1pt}\discretionary{.}{%
}{.}\hspace{.4pt}10\hspace{.1pt}\discretionary{.}{%
}{.}\hspace{.4pt}536208}}


\bibitem{guo2022sd}
Z.~Guo, J.~Tao, S.~Chen, N.~Chawla, and C.~Wang.
\newblock {SD$^{2}$}: Slicing and dicing scholarly data for interactive
  evaluation of academic performance.
\newblock {\em \secondround{IEEE Transactions on Visualization and Computer
  Graphics}}, 2022. \href{https://doi.org/10.1109/TVCG.2022.3163727}
{doi: \textsf{%
10\hspace{.1pt}\discretionary{.}{%
}{.}\hspace{.4pt}1109\discretionary{/}{%
}{/}TVCG\hspace{.1pt}\discretionary{.}{%
}{.}\hspace{.4pt}2022\hspace{.1pt}\discretionary{.}{%
}{.}\hspace{.4pt}3163727}}


\bibitem{hao2022thirty}
H.~Hao, Y.~Cui, Z.~Wang, and Y.-S. Kim.
\newblock Thirty-two years of {IEEE VIS}: Authors, fields of study and
  citations.
\newblock {\em \secondround{IEEE Transactions on Visualization and Computer
  Graphics}}, 29(1):1016--1025, 2022.
  \href{https://doi.org/10.1109/TVCG.2022.3209422}
{doi: \textsf{%
10\hspace{.1pt}\discretionary{.}{%
}{.}\hspace{.4pt}1109\discretionary{/}{%
}{/}TVCG\hspace{.1pt}\discretionary{.}{%
}{.}\hspace{.4pt}2022\hspace{.1pt}\discretionary{.}{%
}{.}\hspace{.4pt}3209422}}


\bibitem{isenberg2016vispubdata}
P.~Isenberg, F.~Heimerl, S.~Koch, T.~Isenberg, P.~Xu, C.~D. Stolper,
  M.~Sedlmair, J.~Chen, T.~M{\"o}ller, and J.~Stasko.
\newblock {Vispubdata.org}: A metadata collection about {IEEE} {Visualization}
  ({VIS}) publications.
\newblock {\em \secondround{IEEE Transactions on Visualization and Computer
  Graphics}}, 23(9):2199--2206, 2016.
  \href{https://doi.org/10.1109/TVCG.2016.2615308}
{doi: \textsf{%
10\hspace{.1pt}\discretionary{.}{%
}{.}\hspace{.4pt}1109\discretionary{/}{%
}{/}TVCG\hspace{.1pt}\discretionary{.}{%
}{.}\hspace{.4pt}2016\hspace{.1pt}\discretionary{.}{%
}{.}\hspace{.4pt}2615308}}


\bibitem{jones2021science}
B.~F. Jones.
\newblock Science and innovation: The under-fueled engine of prosperity.
\newblock {\em Rebuilding the Post-Pandemic Economy, ed. Melissa S. Kearney and
  Amy Ganz (Washington DC: Aspen Institute Press, 2021)}, 2021.

\bibitem{karimi2016inferring}
F.~Karimi, C.~Wagner, F.~Lemmerich, M.~Jadidi, and M.~Strohmaier.
\newblock Inferring gender from names on the web: A comparative evaluation of
  gender detection methods.
\newblock In {\em Proceedings of the International Conference Companion on
  World Wide Web}, pp. 53--54, 2016.
  \href{https://doi.org/10.1145/2872518.2889385}
{doi: \textsf{%
10\hspace{.1pt}\discretionary{.}{%
}{.}\hspace{.4pt}1145\discretionary{/}{%
}{/}2872518\hspace{.1pt}\discretionary{.}{%
}{.}\hspace{.4pt}2889385}}


\bibitem{kim2008visualization}
Y.~G. Kim, J.~H. Suh, and S.~C. Park.
\newblock Visualization of patent analysis for emerging technology.
\newblock {\em Expert Systems with Applications}, 34(3):1804--1812, 2008.
  \href{https://doi.org/10.1016/j.eswa.2007.01.033}
{doi: \textsf{%
10\hspace{.1pt}\discretionary{.}{%
}{.}\hspace{.4pt}1016\discretionary{/}{%
}{/}j\hspace{.1pt}\discretionary{.}{%
}{.}\hspace{.4pt}eswa\hspace{.1pt}\discretionary{.}{%
}{.}\hspace{.4pt}2007\hspace{.1pt}\discretionary{.}{%
}{.}\hspace{.4pt}01\hspace{.1pt}\discretionary{.}{%
}{.}\hspace{.4pt}033}}


\bibitem{kipf2017semi}
T.~N. Kipf and M.~Welling.
\newblock Semi-supervised classification with graph convolutional networks.
\newblock In {\em International Conference on Learning Representations (ICLR)},
  2017. \href{https://doi.org/10.48550/arXiv.1609.02907}
{doi: \textsf{%
10\hspace{.1pt}\discretionary{.}{%
}{.}\hspace{.4pt}48550\discretionary{/}{%
}{/}arXiv\hspace{.1pt}\discretionary{.}{%
}{.}\hspace{.4pt}1609\hspace{.1pt}\discretionary{.}{%
}{.}\hspace{.4pt}02907}}


\bibitem{koch2010iterative}
S.~Koch, H.~Bosch, M.~Giereth, and T.~Ertl.
\newblock Iterative integration of visual insights during scalable patent
  search and analysis.
\newblock {\em \secondround{IEEE Transactions on Visualization and Computer
  Graphics}}, 17(5):557--569, 2010. \href{https://doi.org/10.1109/TVCG.2010.85}
{doi: \textsf{%
10\hspace{.1pt}\discretionary{.}{%
}{.}\hspace{.4pt}1109\discretionary{/}{%
}{/}TVCG\hspace{.1pt}\discretionary{.}{%
}{.}\hspace{.4pt}2010\hspace{.1pt}\discretionary{.}{%
}{.}\hspace{.4pt}85}}


\bibitem{kruskal1983icicle}
J.~B. Kruskal and J.~M. Landwehr.
\newblock Icicle plots: Better displays for hierarchical clustering.
\newblock {\em The American Statistician}, 37(2):162--168, 1983.
  \href{https://doi.org/10.2307/2685881}
{doi: \textsf{%
10\hspace{.1pt}\discretionary{.}{%
}{.}\hspace{.4pt}2307\discretionary{/}{%
}{/}2685881}}


\bibitem{kutz2004examining}
D.~O. Kutz.
\newblock Examining the evolution and distribution of patent classifications.
\newblock In {\em Proceedings of the International Conference on Information
  Visualisation}, pp. 983--988. IEEE, 2004.
  \href{https://doi.org/10.1109/IV.2004.1320261}
{doi: \textsf{%
10\hspace{.1pt}\discretionary{.}{%
}{.}\hspace{.4pt}1109\discretionary{/}{%
}{/}IV\hspace{.1pt}\discretionary{.}{%
}{.}\hspace{.4pt}2004\hspace{.1pt}\discretionary{.}{%
}{.}\hspace{.4pt}1320261}}


\bibitem{latif2018vis}
S.~Latif and F.~Beck.
\newblock {VIS} {Author Profiles}: Interactive descriptions of publication
  records combining text and visualization.
\newblock {\em \secondround{IEEE Transactions on Visualization and Computer
  Graphics}}, 25(1):152--161, 2018.
  \href{https://doi.org/10.1109/TVCG.2018.2865022}
{doi: \textsf{%
10\hspace{.1pt}\discretionary{.}{%
}{.}\hspace{.4pt}1109\discretionary{/}{%
}{/}TVCG\hspace{.1pt}\discretionary{.}{%
}{.}\hspace{.4pt}2018\hspace{.1pt}\discretionary{.}{%
}{.}\hspace{.4pt}2865022}}


\bibitem{li2022gotreescape}
G.~Li and X.~Yuan.
\newblock {GoTreeScape}: Navigate and explore the tree visualization design
  space.
\newblock {\em IEEE Transactions on Visualization and Computer Graphics}, pp.
  1--17, 2022. \href{https://doi.org/10.1109/TVCG.2022.3215070}
{doi: \textsf{%
10\hspace{.1pt}\discretionary{.}{%
}{.}\hspace{.4pt}1109\discretionary{/}{%
}{/}TVCG\hspace{.1pt}\discretionary{.}{%
}{.}\hspace{.4pt}2022\hspace{.1pt}\discretionary{.}{%
}{.}\hspace{.4pt}3215070}}


\bibitem{li2019galex}
Z.~Li, C.~Zhang, S.~Jia, and J.~Zhang.
\newblock Galex: Exploring the evolution and intersection of disciplines.
\newblock {\em \secondround{IEEE Transactions on Visualization and Computer
  Graphics}}, 26(1):1182--1192, 2019.
  \href{https://doi.org/10.1109/TVCG.2019.2934667}
{doi: \textsf{%
10\hspace{.1pt}\discretionary{.}{%
}{.}\hspace{.4pt}1109\discretionary{/}{%
}{/}TVCG\hspace{.1pt}\discretionary{.}{%
}{.}\hspace{.4pt}2019\hspace{.1pt}\discretionary{.}{%
}{.}\hspace{.4pt}2934667}}


\bibitem{liang2022systematic}
W.~Liang, S.~Elrod, D.~A. McFarland, and J.~Zou.
\newblock Systematic analysis of 50 years of {Stanford University} technology
  transfer and commercialization.
\newblock {\em Patterns}, 3(9):100584, 2022.
  \href{https://doi.org/10.1016/j.patter.2022.100584}
{doi: \textsf{%
10\hspace{.1pt}\discretionary{.}{%
}{.}\hspace{.4pt}1016\discretionary{/}{%
}{/}j\hspace{.1pt}\discretionary{.}{%
}{.}\hspace{.4pt}patter\hspace{.1pt}\discretionary{.}{%
}{.}\hspace{.4pt}2022\hspace{.1pt}\discretionary{.}{%
}{.}\hspace{.4pt}100584}}


\bibitem{lin2023sciscinet}
Z.~Lin, Y.~Yin, L.~Liu, and D.~Wang.
\newblock {SciSciNet}: A large-scale open data lake for the science of science
  research.
\newblock {\em Scientific Data}, 10(1):315, 2023.
  \href{https://doi.org/10.1038/s41597-023-02198-9}
{doi: \textsf{%
10\hspace{.1pt}\discretionary{.}{%
}{.}\hspace{.4pt}1038\discretionary{/}{%
}{/}s41597\discretionary{%
}{-}{-}023\discretionary{%
}{-}{-}02198\discretionary{%
}{-}{-}9}}


\bibitem{liu2021understanding}
L.~Liu, N.~Dehmamy, J.~Chown, C.~L. Giles, and D.~Wang.
\newblock Understanding the onset of hot streaks across artistic, cultural, and
  scientific careers.
\newblock {\em Nature Communications}, 12(1):5392, 2021.
  \href{https://doi.org/10.1038/s41467-021-25477-8}
{doi: \textsf{%
10\hspace{.1pt}\discretionary{.}{%
}{.}\hspace{.4pt}1038\discretionary{/}{%
}{/}s41467\discretionary{%
}{-}{-}021\discretionary{%
}{-}{-}25477\discretionary{%
}{-}{-}8}}


\bibitem{marx2020reliance}
M.~Marx and A.~Fuegi.
\newblock Reliance on science: Worldwide front-page patent citations to
  scientific articles.
\newblock {\em Strategic Management Journal}, 41(9):1572--1594, 2020.
  \href{https://doi.org/10.1002/smj.3145}
{doi: \textsf{%
10\hspace{.1pt}\discretionary{.}{%
}{.}\hspace{.4pt}1002\discretionary{/}{%
}{/}smj\hspace{.1pt}\discretionary{.}{%
}{.}\hspace{.4pt}3145}}


\bibitem{marx2022reliance}
M.~Marx and A.~Fuegi.
\newblock Reliance on science by inventors: Hybrid extraction of in-text
  patent-to-article citations.
\newblock {\em Journal of Economics \& Management Strategy}, 31(2):369--392,
  2022. \href{https://doi.org/10.1111/jems.12455}
{doi: \textsf{%
10\hspace{.1pt}\discretionary{.}{%
}{.}\hspace{.4pt}1111\discretionary{/}{%
}{/}jems\hspace{.1pt}\discretionary{.}{%
}{.}\hspace{.4pt}12455}}


\bibitem{morris2002diva}
S.~Morris, C.~DeYong, Z.~Wu, S.~Salman, and D.~Yemenu.
\newblock {DIVA}: A visualization system for exploring document databases for
  technology forecasting.
\newblock {\em Computers \& Industrial Engineering}, 43(4):841--862, 2002.
  \href{https://doi.org/10.1016/S0360-8352(02)00143-2}
{doi: \textsf{%
10\hspace{.1pt}\discretionary{.}{%
}{.}\hspace{.4pt}1016\discretionary{/}{%
}{/}S0360\discretionary{%
}{-}{-}8352\discretionary{%
}{(}{(}02\discretionary{)}{%
}{)}00143\discretionary{%
}{-}{-}2}}


\bibitem{muhlbacher2017treepod}
T.~M{\"u}hlbacher, L.~Linhardt, T.~M{\"o}ller, and H.~Piringer.
\newblock Tree{POD}: Sensitivity-aware selection of {P}areto-optimal decision
  trees.
\newblock {\em \secondround{IEEE Transactions on Visualization and Computer
  Graphics}}, 24(1):174--183, 2017.
  \href{https://doi.org/10.1109/TVCG.2017.2745158}
{doi: \textsf{%
10\hspace{.1pt}\discretionary{.}{%
}{.}\hspace{.4pt}1109\discretionary{/}{%
}{/}TVCG\hspace{.1pt}\discretionary{.}{%
}{.}\hspace{.4pt}2017\hspace{.1pt}\discretionary{.}{%
}{.}\hspace{.4pt}2745158}}


\bibitem{narechania2021vitality}
A.~Narechania, A.~Karduni, R.~Wesslen, and E.~Wall.
\newblock {VITALITY}: Promoting serendipitous discovery of academic literature
  with transformers \& visual analytics.
\newblock {\em \secondround{IEEE Transactions on Visualization and Computer
  Graphics}}, 28(1):486--496, 2021.
  \href{https://doi.org/10.1109/TVCG.2021.3114820}
{doi: \textsf{%
10\hspace{.1pt}\discretionary{.}{%
}{.}\hspace{.4pt}1109\discretionary{/}{%
}{/}TVCG\hspace{.1pt}\discretionary{.}{%
}{.}\hspace{.4pt}2021\hspace{.1pt}\discretionary{.}{%
}{.}\hspace{.4pt}3114820}}


\bibitem{nobre2018juniper}
C.~Nobre, M.~Streit, and A.~Lex.
\newblock Juniper: A tree+ table approach to multivariate graph visualization.
\newblock {\em \secondround{IEEE Transactions on Visualization and Computer
  Graphics}}, 25(1):544--554, 2018.
  \href{https://doi.org/10.1109/TVCG.2018.2865149}
{doi: \textsf{%
10\hspace{.1pt}\discretionary{.}{%
}{.}\hspace{.4pt}1109\discretionary{/}{%
}{/}TVCG\hspace{.1pt}\discretionary{.}{%
}{.}\hspace{.4pt}2018\hspace{.1pt}\discretionary{.}{%
}{.}\hspace{.4pt}2865149}}


\bibitem{pandey2021state}
A.~Pandey, U.~H. Syeda, C.~Shah, J.~A. Guerra-Gomez, and M.~A. Borkin.
\newblock A state-of-the-art survey of tasks for tree design and evaluation
  with a curated task dataset.
\newblock {\em \secondround{IEEE Transactions on Visualization and Computer
  Graphics}}, 28(10):3563--3584, 2021.
  \href{https://doi.org/10.1109/TVCG.2021.3064037}
{doi: \textsf{%
10\hspace{.1pt}\discretionary{.}{%
}{.}\hspace{.4pt}1109\discretionary{/}{%
}{/}TVCG\hspace{.1pt}\discretionary{.}{%
}{.}\hspace{.4pt}2021\hspace{.1pt}\discretionary{.}{%
}{.}\hspace{.4pt}3064037}}


\bibitem{phan2005flow}
D.~Phan, L.~Xiao, R.~Yeh, and P.~Hanrahan.
\newblock Flow map layout.
\newblock In {\em IEEE Symposium on Information Visualization}, pp. 219--224.
  IEEE, 2005. \href{https://doi.org/10.1109/INFVIS.2005.1532150}
{doi: \textsf{%
10\hspace{.1pt}\discretionary{.}{%
}{.}\hspace{.4pt}1109\discretionary{/}{%
}{/}INFVIS\hspace{.1pt}\discretionary{.}{%
}{.}\hspace{.4pt}2005\hspace{.1pt}\discretionary{.}{%
}{.}\hspace{.4pt}1532150}}


\bibitem{qian2021geometric}
Y.~Qian, P.~Expert, P.~Panzarasa, and M.~Barahona.
\newblock Geometric graphs from data to aid classification tasks with {Graph
  Convolutional Networks}.
\newblock {\em Patterns}, 2(4):100237, 2021.
  \href{https://doi.org/10.1016/j.patter.2021.100237}
{doi: \textsf{%
10\hspace{.1pt}\discretionary{.}{%
}{.}\hspace{.4pt}1016\discretionary{/}{%
}{/}j\hspace{.1pt}\discretionary{.}{%
}{.}\hspace{.4pt}patter\hspace{.1pt}\discretionary{.}{%
}{.}\hspace{.4pt}2021\hspace{.1pt}\discretionary{.}{%
}{.}\hspace{.4pt}100237}}


\bibitem{qian2021quantifying}
Y.~Qian, P.~Expert, T.~Rieu, P.~Panzarasa, and M.~Barahona.
\newblock Quantifying the alignment of graph and features in deep learning.
\newblock {\em IEEE Transactions on Neural Networks and Learning Systems},
  33(4):1663--1672, 2021. \href{https://doi.org/10.1109/TNNLS.2020.3043196}
{doi: \textsf{%
10\hspace{.1pt}\discretionary{.}{%
}{.}\hspace{.4pt}1109\discretionary{/}{%
}{/}TNNLS\hspace{.1pt}\discretionary{.}{%
}{.}\hspace{.4pt}2020\hspace{.1pt}\discretionary{.}{%
}{.}\hspace{.4pt}3043196}}


\bibitem{sarvghad2022scientometric}
A.~Sarvghad, R.~Franqui-Nadal, R.~Reznik-Zellen, R.~Chawla, and N.~Mahyar.
\newblock Scientometric analysis of interdisciplinary collaboration and gender
  trends in 30 years of {IEEE VIS} publications.
\newblock {\em \secondround{IEEE Transactions on Visualization and Computer
  Graphics}}, 2022. \href{https://doi.org/10.1109/TVCG.2022.3158236}
{doi: \textsf{%
10\hspace{.1pt}\discretionary{.}{%
}{.}\hspace{.4pt}1109\discretionary{/}{%
}{/}TVCG\hspace{.1pt}\discretionary{.}{%
}{.}\hspace{.4pt}2022\hspace{.1pt}\discretionary{.}{%
}{.}\hspace{.4pt}3158236}}


\bibitem{sedlmair2012design}
M.~Sedlmair, M.~Meyer, and T.~Munzner.
\newblock Design study methodology: Reflections from the trenches and the
  stacks.
\newblock {\em IEEE Transactions on Visualization and Computer Graphics},
  18(12):2431--2440, 2012. \href{https://doi.org/10.1109/TVCG.2012.213}
{doi: \textsf{%
10\hspace{.1pt}\discretionary{.}{%
}{.}\hspace{.4pt}1109\discretionary{/}{%
}{/}TVCG\hspace{.1pt}\discretionary{.}{%
}{.}\hspace{.4pt}2012\hspace{.1pt}\discretionary{.}{%
}{.}\hspace{.4pt}213}}


\bibitem{sun2018effect}
M.~Sun, J.~Zhao, H.~Wu, K.~Luther, C.~North, and N.~Ramakrishnan.
\newblock The effect of edge bundling and seriation on sensemaking of
  biclusters in bipartite graphs.
\newblock {\em \secondround{IEEE Transactions on Visualization and Computer
  Graphics}}, 25(10):2983--2998, 2018.
  \href{https://doi.org/10.1109/TVCG.2018.2861397}
{doi: \textsf{%
10\hspace{.1pt}\discretionary{.}{%
}{.}\hspace{.4pt}1109\discretionary{/}{%
}{/}TVCG\hspace{.1pt}\discretionary{.}{%
}{.}\hspace{.4pt}2018\hspace{.1pt}\discretionary{.}{%
}{.}\hspace{.4pt}2861397}}


\bibitem{tovanich2021gender}
N.~Tovanich, P.~Dragicevic, and P.~Isenberg.
\newblock Gender in 30 years of {IEEE} {Visualization}.
\newblock {\em \secondround{IEEE Transactions on Visualization and Computer
  Graphics}}, 28(1):497--507, 2021.
  \href{https://doi.org/10.1109/TVCG.2021.3114787}
{doi: \textsf{%
10\hspace{.1pt}\discretionary{.}{%
}{.}\hspace{.4pt}1109\discretionary{/}{%
}{/}TVCG\hspace{.1pt}\discretionary{.}{%
}{.}\hspace{.4pt}2021\hspace{.1pt}\discretionary{.}{%
}{.}\hspace{.4pt}3114787}}


\bibitem{uzzi2013atypical}
B.~Uzzi, S.~Mukherjee, M.~Stringer, and B.~Jones.
\newblock Atypical combinations and scientific impact.
\newblock {\em Science}, 342(6157):468--472, 2013.
  \href{https://doi.org/10.1126/science.1240474}
{doi: \textsf{%
10\hspace{.1pt}\discretionary{.}{%
}{.}\hspace{.4pt}1126\discretionary{/}{%
}{/}science\hspace{.1pt}\discretionary{.}{%
}{.}\hspace{.4pt}1240474}}


\bibitem{van2011baobabview}
S.~Van Den~Elzen and J.~J. Van~Wijk.
\newblock {BaobabView}: Interactive construction and analysis of decision
  trees.
\newblock In {\em IEEE Conference on Visual Analytics Science and Technology
  (VAST)}, pp. 151--160. IEEE, 2011.
  \href{https://doi.org/10.1109/VAST.2011.6102453}
{doi: \textsf{%
10\hspace{.1pt}\discretionary{.}{%
}{.}\hspace{.4pt}1109\discretionary{/}{%
}{/}VAST\hspace{.1pt}\discretionary{.}{%
}{.}\hspace{.4pt}2011\hspace{.1pt}\discretionary{.}{%
}{.}\hspace{.4pt}6102453}}


\bibitem{wang2021science}
D.~Wang and A.-L. Barab{\'a}si.
\newblock {\em The science of science}.
\newblock Cambridge University Press, 2021.
  \href{https://doi.org/10.1017/9781108610834}
{doi: \textsf{%
10\hspace{.1pt}\discretionary{.}{%
}{.}\hspace{.4pt}1017\discretionary{/}{%
}{/}9781108610834}}


\bibitem{wang2013quantifying}
D.~Wang, C.~Song, and A.-L. Barab{\'a}si.
\newblock Quantifying long-term scientific impact.
\newblock {\em Science}, 342(6154):127--132, 2013.
  \href{https://doi.org/10.1126/science.1237825}
{doi: \textsf{%
10\hspace{.1pt}\discretionary{.}{%
}{.}\hspace{.4pt}1126\discretionary{/}{%
}{/}science\hspace{.1pt}\discretionary{.}{%
}{.}\hspace{.4pt}1237825}}


\bibitem{wang2019review}
K.~Wang, Z.~Shen, C.~Huang, C.-H. Wu, D.~Eide, Y.~Dong, J.~Qian, A.~Kanakia,
  A.~Chen, and R.~Rogahn.
\newblock A review of {Microsoft Academic Services} for science of science
  studies.
\newblock {\em Frontiers in Big Data}, 2:45, 2019.
  \href{https://doi.org/10.3389/fdata.2019.00045}
{doi: \textsf{%
10\hspace{.1pt}\discretionary{.}{%
}{.}\hspace{.4pt}3389\discretionary{/}{%
}{/}fdata\hspace{.1pt}\discretionary{.}{%
}{.}\hspace{.4pt}2019\hspace{.1pt}\discretionary{.}{%
}{.}\hspace{.4pt}00045}}


\bibitem{wang2021m2lens}
X.~Wang, J.~He, Z.~Jin, M.~Yang, Y.~Wang, and H.~Qu.
\newblock {M2lens}: Visualizing and explaining multimodal models for sentiment
  analysis.
\newblock {\em IEEE Transactions on Visualization and Computer Graphics},
  28(1):802--812, 2021. \href{https://doi.org/10.1109/TVCG.2021.3114794}
{doi: \textsf{%
10\hspace{.1pt}\discretionary{.}{%
}{.}\hspace{.4pt}1109\discretionary{/}{%
}{/}TVCG\hspace{.1pt}\discretionary{.}{%
}{.}\hspace{.4pt}2021\hspace{.1pt}\discretionary{.}{%
}{.}\hspace{.4pt}3114794}}


\bibitem{wang2021interactive}
Y.~Wang, H.~Liang, X.~Shu, J.~Wang, K.~Xu, Z.~Deng, C.~Campbell, B.~Chen,
  Y.~Wu, and H.~Qu.
\newblock Interactive visual exploration of longitudinal historical career
  mobility data.
\newblock {\em IEEE Transactions on Visualization and Computer Graphics},
  28(10):3441--3455, 2021. \href{https://doi.org/10.1109/TVCG.2021.3067200}
{doi: \textsf{%
10\hspace{.1pt}\discretionary{.}{%
}{.}\hspace{.4pt}1109\discretionary{/}{%
}{/}TVCG\hspace{.1pt}\discretionary{.}{%
}{.}\hspace{.4pt}2021\hspace{.1pt}\discretionary{.}{%
}{.}\hspace{.4pt}3067200}}


\bibitem{wang2021seek}
Y.~Wang, T.-Q. Peng, H.~Lu, H.~Wang, X.~Xie, H.~Qu, and Y.~Wu.
\newblock Seek for success: A visualization approach for understanding the
  dynamics of academic careers.
\newblock {\em \secondround{IEEE Transactions on Visualization and Computer
  Graphics}}, 28(1):475--485, 2021.
  \href{https://doi.org/10.1109/TVCG.2021.3114790}
{doi: \textsf{%
10\hspace{.1pt}\discretionary{.}{%
}{.}\hspace{.4pt}1109\discretionary{/}{%
}{/}TVCG\hspace{.1pt}\discretionary{.}{%
}{.}\hspace{.4pt}2021\hspace{.1pt}\discretionary{.}{%
}{.}\hspace{.4pt}3114790}}


\bibitem{wang2019vispubcompas}
Y.~Wang, M.~Yu, G.~Shan, H.-W. Shen, and Z.~Lu.
\newblock {VISPubComPAS}: a comparative analytical system for visualization
  publication data.
\newblock {\em Journal of Visualization}, 22:941--953, 2019.
  \href{https://doi.org/10.1007/s12650-019-00585-2}
{doi: \textsf{%
10\hspace{.1pt}\discretionary{.}{%
}{.}\hspace{.4pt}1007\discretionary{/}{%
}{/}s12650\discretionary{%
}{-}{-}019\discretionary{%
}{-}{-}00585\discretionary{%
}{-}{-}2}}


\bibitem{windhager2015concept}
F.~Windhager, A.~Amor-Amor{\'o}s, M.~Smuc, P.~Federico, L.~ZenkVxInsight, and
  S.~Miksch.
\newblock A concept for the exploratory visualization of patent network
  dynamics.
\newblock In {\em IVAPP}, pp. 268--273, 2015.
  \href{https://doi.org/10.5220/0005360002680273}
{doi: \textsf{%
10\hspace{.1pt}\discretionary{.}{%
}{.}\hspace{.4pt}5220\discretionary{/}{%
}{/}0005360002680273}}


\bibitem{wu2022defence}
A.~Wu, D.~Deng, F.~Cheng, Y.~Wu, S.~Liu, and H.~Qu.
\newblock In defence of visual analytics systems: Replies to critics.
\newblock {\em \secondround{IEEE Transactions on Visualization and Computer
  Graphics}}, 29(1):1026--1036, 2022.
  \href{https://doi.org/10.1109/TVCG.2022.3209360}
{doi: \textsf{%
10\hspace{.1pt}\discretionary{.}{%
}{.}\hspace{.4pt}1109\discretionary{/}{%
}{/}TVCG\hspace{.1pt}\discretionary{.}{%
}{.}\hspace{.4pt}2022\hspace{.1pt}\discretionary{.}{%
}{.}\hspace{.4pt}3209360}}


\bibitem{wu2019large}
L.~Wu, D.~Wang, and J.~A. Evans.
\newblock Large teams develop and small teams disrupt science and technology.
\newblock {\em Nature}, 566(7744):378--382, 2019.
  \href{https://doi.org/10.1038/s41586-019-0941-9}
{doi: \textsf{%
10\hspace{.1pt}\discretionary{.}{%
}{.}\hspace{.4pt}1038\discretionary{/}{%
}{/}s41586\discretionary{%
}{-}{-}019\discretionary{%
}{-}{-}0941\discretionary{%
}{-}{-}9}}


\bibitem{wu2015egoslider}
Y.~Wu, N.~Pitipornvivat, J.~Zhao, S.~Yang, G.~Huang, and H.~Qu.
\newblock {egoSlider}: Visual analysis of egocentric network evolution.
\newblock {\em \secondround{IEEE Transactions on Visualization and Computer
  Graphics}}, 22(1):260--269, 2015.
  \href{https://doi.org/10.1109/TVCG.2015.2468151}
{doi: \textsf{%
10\hspace{.1pt}\discretionary{.}{%
}{.}\hspace{.4pt}1109\discretionary{/}{%
}{/}TVCG\hspace{.1pt}\discretionary{.}{%
}{.}\hspace{.4pt}2015\hspace{.1pt}\discretionary{.}{%
}{.}\hspace{.4pt}2468151}}


\bibitem{ye2022visatlas}
Y.~Ye, R.~Huang, and W.~Zeng.
\newblock {VISAtlas}: An image-based exploration and query system for large
  visualization collections via neural image embedding.
\newblock {\em IEEE Transactions on Visualization and Computer Graphics}, pp.
  1--15, 2022. \href{https://doi.org/10.1109/TVCG.2022.3229023}
{doi: \textsf{%
10\hspace{.1pt}\discretionary{.}{%
}{.}\hspace{.4pt}1109\discretionary{/}{%
}{/}TVCG\hspace{.1pt}\discretionary{.}{%
}{.}\hspace{.4pt}2022\hspace{.1pt}\discretionary{.}{%
}{.}\hspace{.4pt}3229023}}


\bibitem{yin2022public}
Y.~Yin, Y.~Dong, K.~Wang, D.~Wang, and B.~F. Jones.
\newblock Public use and public funding of science.
\newblock {\em Nature Human Behaviour}, 6(10):1344--1350, 2022.
  \href{https://doi.org/10.1038/s41562-022-01397-5}
{doi: \textsf{%
10\hspace{.1pt}\discretionary{.}{%
}{.}\hspace{.4pt}1038\discretionary{/}{%
}{/}s41562\discretionary{%
}{-}{-}022\discretionary{%
}{-}{-}01397\discretionary{%
}{-}{-}5}}


\bibitem{yin2021coevolution}
Y.~Yin, J.~Gao, B.~F. Jones, and D.~Wang.
\newblock Coevolution of policy and science during the pandemic.
\newblock {\em Science}, 371(6525):128--130, 2021.
  \href{https://doi.org/10.1126/science.abe3084}
{doi: \textsf{%
10\hspace{.1pt}\discretionary{.}{%
}{.}\hspace{.4pt}1126\discretionary{/}{%
}{/}science\hspace{.1pt}\discretionary{.}{%
}{.}\hspace{.4pt}abe3084}}


\end{thebibliography}

\end{document}